\documentclass[12pt]{article} 
\usepackage{graphicx} 
\usepackage{amssymb} 
\usepackage{amsmath}

\newcommand{\sechead}[1]{\par\vspace{0.25in}{\bf #1}\nopagebreak\vspace{0.25in}\nopagebreak}

\begin{document} 
\begin{titlepage}
	
	\vskip 2cm 
	\begin{center}
		\large{{\bf From Surface Operators to Non-Abelian \\
		Volume Operators in Puff Field Theory}} 
	\end{center}
	
	\vskip 2cm 
	\begin{center}
		Vatche Sahakian\footnote{\texttt{sahakian@theory.caltech.edu}} \\
		\vskip 24pt {\sl Harvey Mudd College, Claremont, CA 91711 USA} \\
		{\sl California Institute of Technology, Pasadena, CA 91125, USA} 
	\end{center}
	
	\vskip 2cm 
	\begin{abstract}
		Puff Field Theory is a low energy decoupling regime of string theory that still retains the non-local attributes of the parent theory - while preserving isotropy for its non-local degrees of freedom. It realizes an extended holographic dictionary at strong coupling and dynamical non-local states akin to defects or the surface operators of local gauge theories. In this work, we probe the non-local features of PFT using D3 branes. We find supersymmetric configurations that end on defects endowed with non-Abelian degrees of freedom. These are $2+1$ dimensional defects in the $3+1$ dimensional PFT that may be viewed as volume operators. We determine their R-charge, vacuum expectation value, energy, and gauge group structure.
	\end{abstract}
\end{titlepage}

\newpage \setcounter{page}{1} 

\section{Introduction and Results}

Surface operators or defects~\cite{Gukov:2006jk}-\cite{Gukov:2008sn} - higher dimensional generalizations of Wilson and 't Hooft loops~\cite{Rey:1998ik}-\cite{Chen:2006iu} - are interesting non-local probes of gauge theories. Beside their underlying rich mathematical structure, they encode physical information about the parent theory they are inserted into through various embedding-related consistency conditions. From the string theory perspective, they may be related to intersecting brane constructions and hence probe string and M theory directly. 

In certain low energy scaling regimes, string theory is known to admit interesting gravity-decoupled settings that retain some of the non-local attributes of the parent theory~\cite{Seiberg:1999vs}. These come in various flavors, from non-commutative field theories~\cite{Hashimoto:1999ut,Maldacena:1999mh,Cai:1999aw} to theories of open strings~\cite{Klebanov:2000pp,Seiberg:2000ms,Harmark:2000wv} and membranes~\cite{Gopakumar:2000ep} and dipoles~\cite{Bergman:2000cw}-\cite{Alishahiha:2002ex}. In previous works, Wilson lines have been shown to play a particularly
important role in understanding novel non-local gauge invariant features of such theories~\cite{Ishibashi:1999hs,Gross:2000ba,Rozali:2000np}. The purpose of this work is to use surface operators to explore a particular scaling regime of string theory known as Puff Field Theory~\cite{Ganor:2006ub}-\cite{FreedBrown:2009py}. 

Puff Field Theory (PFT) lies in a large class of non-local theories that can be constructed by considering D-branes in Melvin backgrounds. Starting with flat space with metric written in cylindrical coordinates
\begin{equation}
	ds^2 = dz^2 + dr^2 + r^2\,d\phi^2\ ,
\end{equation}
Melvin geometry involves a simultaneous twist of the form~\cite{Ganor:2006ub,Dudas:2001ux,Dhokarh:2007ry,Dhokarh:2008ki}
\begin{equation}\label{eq:twist}
	z\simeq z + 2\,\pi\, R\ \ \ \mbox{and}\ \ \ \phi \simeq \phi + 2\,\pi \eta R
\end{equation}
where $\eta$ is the twist parameter. Alternatively, we can write the metric as
\begin{equation}
	ds^2 = dz^2 + dr^2 + r^2 (d\phi+\eta\, dz)^2
\end{equation}
with $\phi\sim \phi+2\pi$, and $z\sim z+2\,\pi\,R$. In general, arranging D-branes in Melvin backgrounds characteristically leads to non-local worldvolume theories. Heuristically, one can think of the origin of the non-locality as arising from open strings whose endpoints are spread out due to the twist - or equivalently by a polarizing flux~\cite{Bigatti:1999iz,Seiberg:2000ms}. \cite{Bergman:2000cw,Bergman:2001rw,Hashimoto:2004pb,Hashimoto:2005hy,Lunin:2005jy} catalogue various possibilities - depending on how D-branes are arranged with respect to the twist - and demonstrate that this construction leads to theories related to non-commutative gauge theories and Non-Commutative Open String theory.  

PFT was introduced in~\cite{Ganor:2006ub} through a similar setup - constructed from D0 branes in a Melvin background. PFTs can come in many flavors, differing in worldvolume dimensionality and amount of supersymmetry. The PFT of interest in this work can be defined as follows: start with a Melvin background in M theory with $N$ units of momentum along $z$; reduce along $z$ to IIA theory and D0 branes in a Melvin universe with electric {\em and} magnetic RR flux; and T-dualize along transverse directions to $z$, $r$, and $\phi$. With three T-dualities and a proper decoupling limit, we get to $3+1$ dimensional PFT - the worldvolume theory of $N$ D3 branes in a Melvin universe. One can see hints of non-local dynamics in the decoupled theory as follows: a mode with $j$ units of angular momentum along $\phi$ translates to a fractional D3 brane charge of $\nu=j \eta R$. Such states are expected to occupy a volume proportional to $\nu$, as if we have a D3 brane `puff' or bubble on the PFT worldvolume. Unlike other non-local theories mentioned earlier, it is proposed that PFT realizes $3+1$ dimensional non-locality while preserving full rotational $SO(3)$ symmetry~\cite{Ganor:2006ub}. This is easiest to see from the holographic dual geometry that we will study in this work later on. Such a setting is of particular interest as it lends itself to cosmological applications. In~\cite{Minton:2007fd} for example, PFT was used to model a strongly coupled non-local primordial plasma and compute signatures of non-locality in the Cosmic Microwave Background radiation. 

In this work, we consider a particular realization of a PFT arising from D3 branes in a Melvin universe with a slightly more elaborate structure involving two angular twists. The details of this construction can be found in~\cite{Ganor:2007qh}. The end result is: $\mathcal{N}=2$, $3+1$ dimensional PFT with $U(1)\times U(2)$ R-symmetry. And the non-local states in this theory carry R-charge inside the $U(2)$. 

Unfortunately, PFTs are still not well understood. In~\cite{Haque:2008vm}, Morita equivalence~\cite{Seiberg:1999vs,Schwarz:1998qj,Hashimoto:1999yj,Pioline:1999xg} was used to relate $0+1$ dimensional PFT  - with rational twist parameter $\eta R=-r/s$ - to a theory holographically dual to $AdS_5\times S^5/Z_s$ with electric and magnetic RR fluxes. It was proposed through a chain of duality transformations that this system is related to $2+1$ dimensional SYM with a 't Hooft flux, matter in the fundamental, and twisted flavor. A top-down treatment akin to~\cite{Seiberg:1999vs} by considering open strings in the Melvin background is also a difficult task that has yet not been completed\footnote{See however~\cite{Dudas:2001ux,Dhokarh:2007ry} for treatments in somewhat different but related settings.}. On the other hand, one has decent computational handle on the strong coupling regime of the theory through the holographic dual geometry constructed in~\cite{Ganor:2007qh}. 
\begin{figure}
	\begin{center}
		\includegraphics[width=6in]{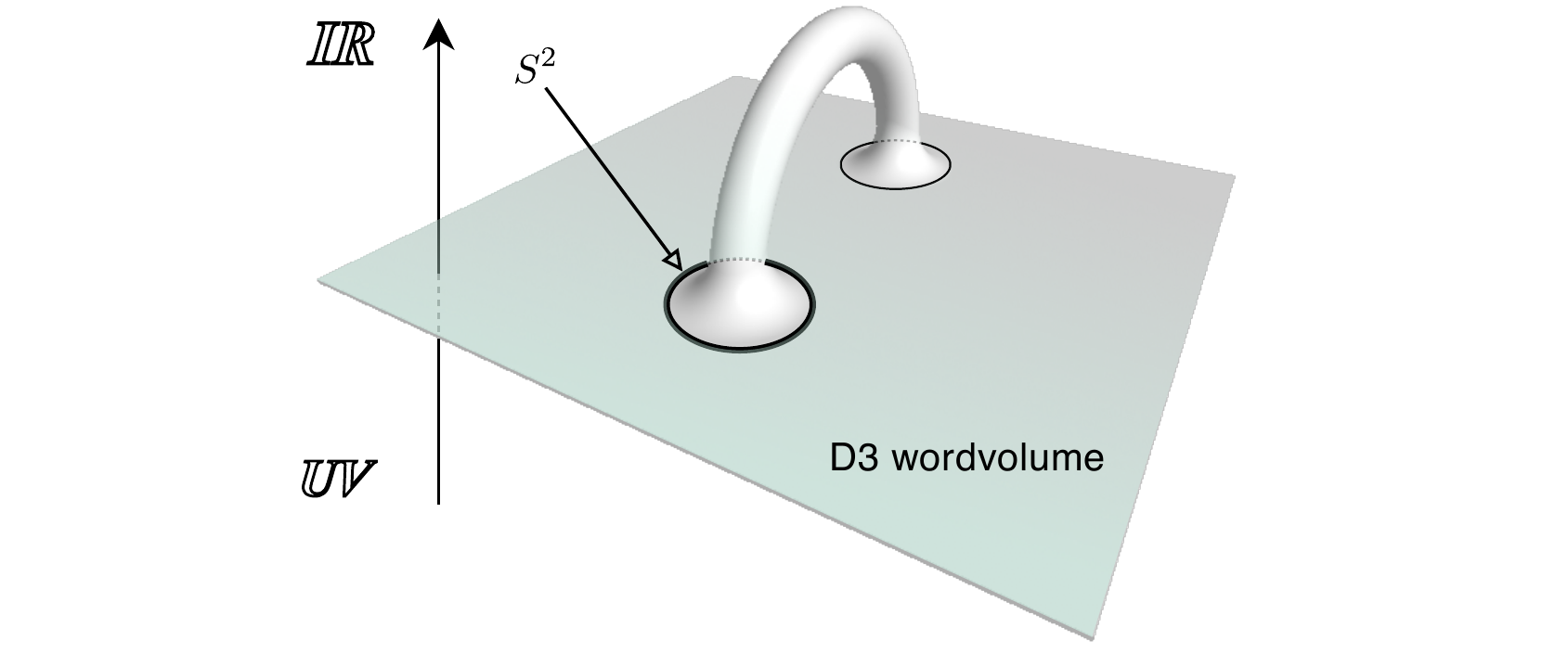} 
	\end{center}
	\caption{\em A cartoon of how a correlator of two non-local operator insertions in the PFT would look like in the holographic dual picture.} \label{fig:tentacle} 
\end{figure}
In~\cite{FreedBrown:2009py}, this geometry was probed using geodesics and evidence was presented that the PFT puffs may be the footprints of D3 brane tentacles or protrusions inserted onto the PFT worldvolume, as cartooned in Figure~\ref{fig:tentacle}. Non-local operator insertions get inserted in the bulk instead of the boundary - at an extent related to their puff size as determined by the UV-IR correspondence. And the bulk space has two sectors, one for describing puff dynamics at wavelengths larger than their size, and another for the internal degrees of freedom of puffs. The two sectors also appear quantum entangled~\cite{FreedBrown:2009py}. In general, the concluding picture shown in Figure~\ref{fig:tentacle} is obviously reminiscent of defects arising in local gauge theories.
In this work, we want to continue probing PFT at strong coupling using gravitational holography~\cite{Maldacena:1997re,Witten:1998qj,Gubser:1998bc} - this time using D-brane configurations similar to the ordered surface operators of local gauge theories. The picture of the non-local puffs developed in~\cite{FreedBrown:2009py} serves as an inspiration to this exercise.

Hence, the question we want to answer is the following: what are all 1/2 BPS surface operators of $\mathcal{N}=2$ $3+1$ dimensional PFT, realized as D3 brane probes of the holographic dual geometry~\cite{Buchbinder:2007ar,Drukker:2008wr,Koh:2008kt}. To be more careful with the nomenclature, we seek defects extended in two spatial directions of the $3+1$ dimensional PFT; strictly speaking, we should call these `volume operators'. Unlike the usual surface operator program, in this case we know little about the PFT field content. We have no equivalent to the Hitchin equation~\cite{Gukov:2006jk,Kapustin:2006pk}, but we can study supersymmetric embedding of D3 branes in the PFT holographic background within a setup that lends itself naturally to a surface/volume operator interpretation: D3 branes which land at the UV boundary of the bulk space onto a two dimensional planar configuration and leave 4 supersymmetries unbroken. From the perspective of the decoupled PFT worldvolume theory, these insertions will look like co-dimension one defects. We then expect that these volume operators would be of the order type, \`{a} la Wilson operators, as opposed to the 't Hooft-like disorder type associated with singularities in the worldvolume fields. We can compute the vacuum expectation value, energy, and R-charge of these configurations at strong coupling using the bulk dynamics. We would hope that this data will help in understanding or deconstructing the PFT.

Before we delve into the computational details, we summarize our results which can be neatly collected in the table below.
\begin{center}
	\begin{tabular}
		{|c||ccccc|} \hline \textbf{\em Section} & BPS condition & Type & AdS? & Parameters & Gauge Group\\
		\hline \hline \ref{sub:scenarioI} & $(1\pm\gamma^{05})\epsilon_0=0$ & Waves & Yes & 4 functions & \\
		\hline \ref{subsub:scenarioIIa} & $(1+i \gamma^{0513})\epsilon_0=0$ & Clover VO & No & 4 constants & $U(8)$\\
		\hline \ref{subsub:scenarioIIbc} & $(1-i \gamma^{0512})\epsilon_0=0$ & Figure Eight VO & No & 4 constants & $U(2)\times U(2)$\\
		\hline \ref{subsub:scenarioIIbc} & $(1+i \gamma^{0518})\epsilon_0=0$ & Figure Eight VO & No & 4 constants & $U(2)\times U(2)$ \\
		\hline 
	\end{tabular}
\end{center}
The table lists four 1/2 BPS configurations of which three are candidate volume operators (VO) of PFT. The first column indicates the section of the text where the corresponding discussion can be found. The second column shows the BPS condition for the associated operator. $\epsilon_0$ is a complex chiral spinor of IIB supergravity. And the indices on the gamma matrices are as follows: $0$ is the time direction; $5$ is the direction along the worldvolume of the PFT but transverse to the defect (the defect extends in the $6$ and $7$ directions along the worldvolume); $1$ is the Melvin twist direction; $3$ is the holographic direction transverse to the PFT worldvolume; and $2$ and $8$ parametrize a transverse 2-sphere that serves as a base over which the Melvin twist coordinate $1$ is fibered. The case in the first row corresponds to simply adding a probe with a plane wave that breaks an additional 50\% of the supersymmetries and hence is of no interest to us. The fourth column labeled `AdS' lists whether the configuration is BPS for the $AdS_5\times S^5$ background - confirming that all volume operators we have found are proper to the PFT. The fifth column lists the numbers of free parameters or functions for each configuration - hence none of the volume operators are rigid. And the last column lists the maximal gauge group that the degrees of freedom of the corresponding defect can realize.

\begin{figure}
	\begin{center}
		\includegraphics[width=7in]{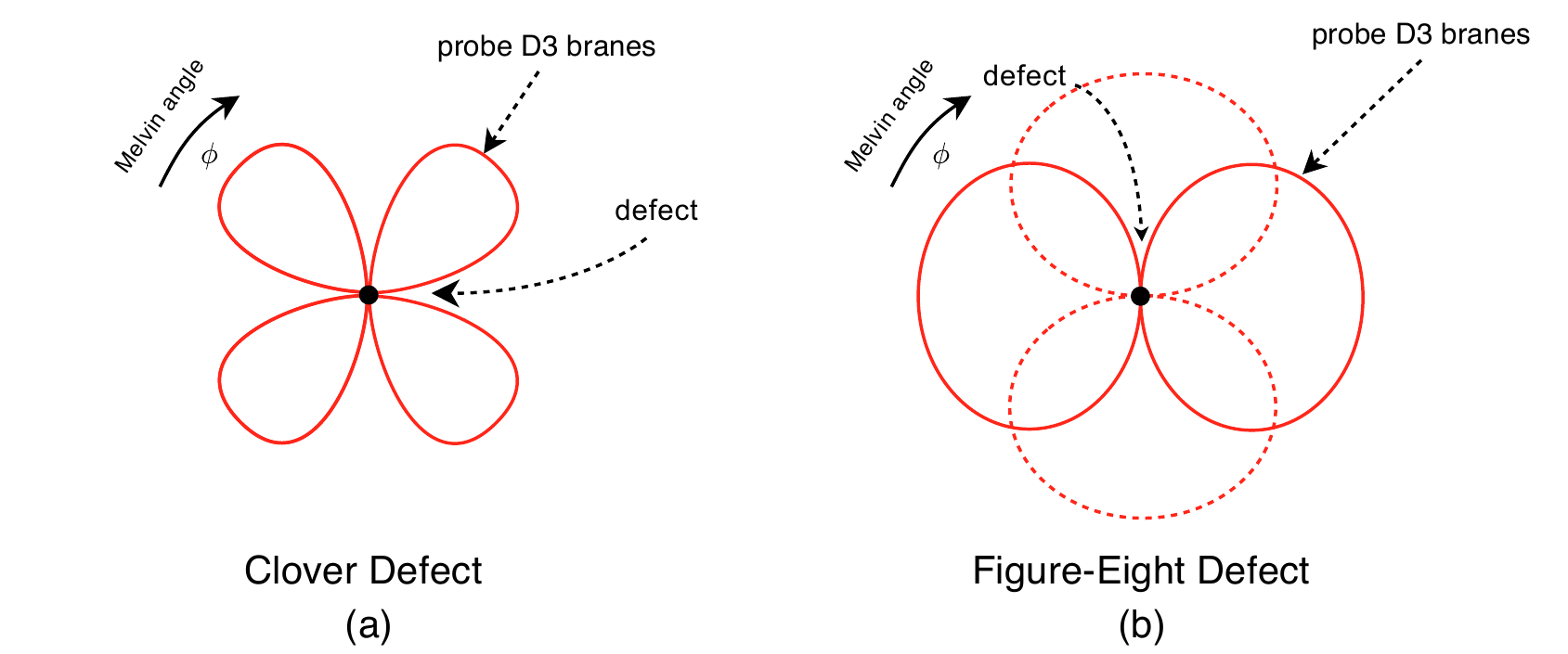} 
	\end{center}
	\caption{\em (a) The clover defect of Section~\ref{subsub:scenarioIIa}. The defect is a spatially two-dimensional planar configuration on the worldvolume of the source branes. The angular direction shown represents the Melvin twist angle $\phi$. (b) A depiction of the figure eight defect of Section~\ref{subsub:scenarioIIbc}. In this case, we have half as many branches of the probe brane landing on the defect. There are two possible configurations shown separately with solid and dashed lines differing also in coordinate embeddings not shown. Each breaks a different set of supersymmetries, shown in the third and fourth row of the table in the text. Each of (a) and (b) has four parameters, two of which are most apparent in this figure: (1) a shift in $\phi$, and (2) the maximum distance in the bulk away from the defect the probes extend.} \label{fig:main} 
\end{figure}

Our volume operators exhibit an interesting new structure depicted in Figure~\ref{fig:main}. The Figure shows a sectional embedding of the probe D3 branes: (a) shows the `clover' defect of Section \ref{subsub:scenarioIIa}, (b) shows the `figure eight' defects of Section \ref{subsub:scenarioIIbc}.
The radial direction in the figure is the transverse holographic coordinate with UV being at the defect and IR being away from it; while the angular direction is the Melvin twist angle. The D3 probes have two of their three spatial directions fixed onto the defects, while the third expands transversely to the PFT worldvolume - with a particularly interesting profile in the Melvin twist direction.
In all cases, the probe brane folds in the bulk, turns around and lands back at the UV boundary onto the defect, enhancing the defect degrees of freedom with non-Abelian structure. In the first case, Figure~\ref{fig:main}(a), the `clover defect' is the footprint of four branches of the probe. Each branch can yield a $U(2)$ gauge group on the defect. In the UV, the four branches of the D3 branes may be left with enough massless degrees of freedom to yield $U(8)$ structure; or break the group all the way down to $U(2)\times U(2)\times U(2)\times U(2)$. Figure~\ref{fig:main}(b) on the other hand shows the `figure eight' defects which have two branches. They can realize $U(2)\times U(2)$ gauge group only. The configuration comes in two forms, shown with a solid and dashed line, corresponding to the third and fourth row in the table above. The two differ in their embedding in the other transverse directions.

In all cases, the volume operators have non-trivial vacuum expectation values similar to Wilson loops - but instead of an exponential of the area of the probe, one has the exponential of the probe volume {\em per puff volume}. The R-charge is also non-zero. This strongly suggests that the non-local states of PFT should be view as D3 brane protrusions from the worldvolume. These D3 brane configurations are {\em not} static. At the UV boundary, the defects or footprints of the probes are fixed in time and space from the perspective of the PFT theory. But in the bulk, the profile evolve in time in interesting ways while remaining attached to the fixed defect at the boundary. The holographic bulk of the PFT is a spinning black hole; it is not surprising that probes in this background would be non-static. Indeed, we also show that the PFT holographic background does not admit any non-trivial static supersymmetric D3 brane configurations with co-dimension one boundary in the UV. Many of these findings are new to surface/volume operators as well as non-local field theories in general.

The outline of the paper goes as follows. In Section 2, we set up the problem and review the PFT holographic geometry. In Section 3, we present the general treatment for solving the BPS conditions for D3 brane probes in the given background. Section 4 collects all the detailed results. Section 5 presents some speculations on how to realize spherical defects instead of planar ones. Section 6 discusses the significance of the results and future directions. Appendix A collects the gory details about spinors and BPS equations. 

\section{Setup} \label{sec:equations} 

Our goal is to find $1/2$ BPS volume operators in $3+1$ dimensional strongly coupled $\mathcal{N}=2$ PFT. We start with the holographic dual background geometry introduced in~\cite{Ganor:2007qh}, and look for supersymmetric probe D3 brane embeddings in this background. We first need to find the Killing spinors for the geometry using~\cite{Schwarz:1983qr}\cite{Howe:1983sr}\cite{Green:1998by}
\begin{equation}
	\label{eq:killing} {
	\partial_\mu\epsilon} + \frac{1}{4}\, \omega_{\mu,{ab}} \gamma^{ab}\epsilon =\frac{i}{1920} {\mathcal{F}_{\rho_1 \rho_2 \rho_3 \rho_4 \rho_5} \gamma^{\rho_1 \rho_2 \rho_3 \rho_4 \rho_5}\gamma_{\mu }\epsilon} 
\end{equation}
where $\mathcal{F}^{(5)}$ is the RR five-form field strength of IIB supergravity and $\omega^{(1)}$ is the gravitational connection. We need to solve for $\epsilon$, a complex chiral spinor. Our conventions are such that the supergravity equations of motion for the relevant sector of the background fields look like~\cite{Howe:1983sr}
\begin{equation}
	\label{eq:bgeom} R_{\mu \nu }=\frac{1}{96} \mathcal{F}_{\rho_1 \rho_2 \rho_3 \rho_4 \mu} \mathcal{F}^{\rho_1 \rho_2 \rho_3 \rho_4}_{\ \ \ \ \ \ \ \ \ \nu }\ . 
\end{equation}

Given $\epsilon$, the task is then to find D3 brane embeddings that solve the worldvolume BPS condition arising from $\kappa$-symmetry~\cite{Aganagic:1996nn}\cite{Cederwall:1996pv}
\begin{equation}
	\label{eq:bps} \epsilon -\frac{i}{\mathcal{L}_{DBI}}{ 
	\partial_{\sigma^0} x^{\mu_1} 
	\partial_{\sigma^1} x^{\mu_2} 
	\partial_{\sigma^2} x^{\mu_3} 
	\partial_{\sigma^3} x^{\mu_4} \gamma_{\mu_1 \mu_2 \mu_3 \mu_4}\epsilon}=0 
\end{equation}
where the $\sigma^i$'s are worldvolume coordinates and the $x^\mu(\sigma)$'s are the target space embedding functions to solve for. We use the static gauge throughout the computations. The D3 brane Lagrangian is given by \label{sec:lagrangian} 
\begin{equation}
	\label{eq:action} \mathcal{L} = \mathcal{L}_{DBI}+\mathcal{L}_{WZ} 
\end{equation}
with 
\begin{equation}
	\label{eq:dbi} \mathcal{L}_{DBI} = -\mathcal{T}\,\sqrt{-\mbox{Det} \left[ g_{\mu\nu} 
	\partial_a x^\mu 
	\partial_b x^\nu \right] } 
\end{equation}
and the Wess-Zumino term 
\begin{equation}
	\label{eq:wz} \mathcal{L}_{WZ} = \mathcal{T}\,\mathcal{C}_{\rho_1\rho_2\rho_3\rho_4} 
	\partial_0 x^{\rho_1}\, 
	\partial_1 x^{\rho_2}\, 
	\partial_2 x^{\rho_3}\, 
	\partial_3 x^{\rho_4} 
\end{equation}
with $\mathcal{F}^{(5)} = d \mathcal{C}^{(4)}$. In our conventions, the tension of the D3 brane would be written as 
\begin{equation}
	\mathcal{T}=\frac{1}{(2 \pi)^3 g_s {\alpha'}^2} 
\end{equation}
where $g_s$ is the IIB string coupling. We need~(\ref{eq:action}) when computing the action evaluated at a BPS configuration, as well as when determining the energy and R-charges. We start by reviewing the holographic background geometry of interest.

\subsection{Background geometry} 

\label{sec:background_fields}

The background geometry is that of a large number of D3 branes in a Melvin universe, constructed in detail in~\cite{Ganor:2007qh}. We label the coordinates as 
\begin{equation}
	\{T,\phi ,\varphi , \theta, w,\psi , X^1, X^2, X^3 ,\chi \} 
\end{equation}
where the source D3 branes extend in the $X^i$ directions; $w$ is the holographic coordinate proportional to PFT energy scale in the UV-IR correspondence: large $w$ corresponds to the UV, small $w$ to the IR. The remaining angular coordinates $\phi ,\varphi ,\theta,\chi$ carry the $U(1)\times U(2)$ isometries associated with the R-symmetry group of the PFT. $\psi$ plays a special role in the holographic dictionary as we will see later. These coordinates are bounded as follows 
\begin{equation}
	0\leq\psi\leq\pi/2\ \ ,\ \ 0\leq\theta\leq\pi\ \ ,\ \ 0\leq\varphi\leq 2\,\pi\ \ ,\ \ 0\leq\phi\leq 2\,\pi \ \ ,\ \ 0\leq\chi\leq 2\,\pi 
\end{equation}
where $\theta$, $\varphi$, and $\phi$ are derived from Hopf fibration coordinates on a 3-sphere - with $\phi$ being the fiber - which, for the topology at hand, is a non-contractible cycle over the $S^2$ base parameterized by $\theta$ and $\varphi$. The Melvin twist underlying the construction of the background can be traced to  the $\phi$ coordinate. The string frame metric is given by the spinning geometry~\cite{Ganor:2007qh}
\begin{eqnarray}
	\label{eq:metric} &&{ds_{str}^2}= \alpha' \sqrt{G}\, H \left[ \frac{dw^2}{w^2} - w^2 dT^2 +d\psi^2 + \frac{\cos^2\psi}{4} \left( d\theta^2+\sin^2\theta d\varphi^2 \right) + \sin^2\psi\, d\chi^2 \right] \nonumber \\
	&&+ \frac{\alpha' \sqrt{G}}{H} \left[ w^2 dX_i^2 + \cos^2\psi \left( d\phi - f(\theta) d\varphi + w^4 dT \right)^2 \right] 
\end{eqnarray}
where 
\begin{equation}
	H\equiv \sqrt{1+w^6 \cos^2 \psi}\ \ \ ,\ \ \ f(\theta )\equiv \frac{1}{2} (1-\cos \theta)\ . 
\end{equation}
The dilaton is constant 
\begin{equation}
	e^\Phi = g_s 
\end{equation}
and the parameter $G$ is given by 
\begin{equation}
	G\equiv 4\pi g_s N \ . 
\end{equation}
with $N$ being the number of source D3 branes. $G$ plays the role of coupling constant in the strongly coupled PFT - at large $N$. The RR 4-form gauge field is given by 
\begin{equation}
	\mathcal{C}_{{TX_1 X_2 X3}}=\frac{1}{H^2} \ \ , \ \ \mathcal{C}_{{\phi X_1 X_2 X3}}=-\frac{ w^2 \cos^2 \psi}{H^2} \ \ , \ \ \mathcal{C}_{{\varphi X_1 X_2 X3}}=\frac{ w^2 f(\theta ) \cos^2 \psi}{H^2} 
\end{equation}
with all other components zero. All of our coordinates are dimensionless, scaled to absorb the physical parameters of the problem. In particular, the time coordinate $T$ and the $X^i$'s are related to the physical coordinates of the dual PFT theory $t$ and $x^i$ by\footnote{To relate our coordinates to ones appearing in the literature, we have $z=1/V$ and $\xi=1/w$ to relate to~\cite{FreedBrown:2009py}. And $w\equiv {V\,\Delta}/{G^{1/6}}$, $X_i\equiv {x_i}/{G^{1/3} \Delta}$, $T\equiv {t}/{G^{1/3} \Delta}$ to relate to~\cite{Ganor:2007qh}. } 
\begin{equation}
	X^i\equiv \frac{x^i}{G^{1/3} \Delta}\ \ \ ,\ \ \ T \equiv \frac{t}{G^{1/3} \Delta}\ . 
\end{equation}
$\Delta$ (which has units of length) sets the scale of non-locality in the dual PFT (related to $\eta$ in the Introduction). At large $N$ and strong coupling, this scale comes dressed with a factor of $G^{1/3}$ as seen from these expressions. Looking back at the background fields in these dimensionless coordinates, we see that our system has two effective parameters in the large $N$ limit and at strong coupling: 
\begin{equation}
	G \ \ \ \mbox{and}\ \ \ G^{1/3} \Delta 
\end{equation}
{\em i.e.} a coupling constant and a scale of non-locality. The ten dimensional gravitational coupling is given by $\kappa_{10} \simeq g_s {\alpha'} ^2$ and the metric~(\ref{eq:metric}) describes the geometry {\em after} taking the decoupling limit $\alpha'\rightarrow 0$. This requires holding $g_{YM}^2$ and $\Delta^3=\eta \alpha^2$ fixed, where $\eta$ is the twist parameter introduced in equation~(\ref{eq:twist}).  The AdS limit of this background geometry can be reached by simply taking $w^2\rightarrow 0\Rightarrow H\rightarrow 1$. 

We also need the vielbein. The diagonal part of the metric is easy to handle. The subspace spanned by $T$, $\phi$, and $\varphi$ is however trickier. We will choose an upper triangular gauge and write 
\begin{equation}
	e^a_{\ \mu }=\left( 
	\begin{array}{ccc}
		\frac{w}{\sqrt{H}} & -\frac{w^3 \cos^2 \psi}{\sqrt{H}} & \frac{w^3 \cos^2 \psi f(\theta )}{\sqrt{H}} \\
		0 & \cos \psi \sqrt{H} & -\cos \psi f(\theta ) \sqrt{H} \\
		0 & 0 & \frac{1}{2}\cos \psi \sin\theta \sqrt{H} 
	\end{array}
	\right) 
\end{equation}
where we write only the $T$, $\phi$, and $\varphi$ subspace; the rest of the vielbein is diagonal. This choice makes the computations considerably more tractable. 

\sechead{Regime of validity}

The background geometry is not reliable everywhere. In particular, the metric has a singularity at $\psi=\pi/2$. We restrict our computations to the $\psi=0$ plane throughout; and in this regime, for small curvature scales compared to the string scale, we need the generic strong coupling condition
\begin{equation}
	\label{eq:strongcoupling} G \gg 1\ . 
\end{equation}
To avoid strings wrapping the $\phi$ direction becoming too light, we need 
\begin{equation}
	\label{eq:tduality} \frac{\sqrt{G} \cos^2\psi}{H} \gg 1 
\end{equation}
which adds an upper UV bound on $w$ 
\begin{equation}
	\label{eq:conditionxi} w \ll G^{1/6} \ . 
\end{equation}
Otherwise, we would need to consider the T-dual geometry. This leads to IIA theory with D4 branes. Pushing the dual circle to larger sizes lifts the picture to M-theory and M5 branes\footnote{In these dual pictures, the planar defects we consider in this work correspond to probe D4 branes and probe D2 branes. In the M-theory picture, we would have probe M5 and M2 branes with boundaries on an M5 brane. The theory has a consistent UV completion in M-theory and physical observables - such as the value of the action of a probe - would be invariants under these duality frame changes. In the extreme UV, the M-theory holographic background becomes parametrically flatter with higher energy.}.

Finally, we also require weak string coupling 
\begin{equation}
	\label{eq:weakstringcoupling} g_s= \frac{G}{4\pi N}\ll 1 \ . 
\end{equation}
Using~(\ref{eq:strongcoupling}), this implies that one needs $N\gg 1$. In conclusion, by making $G$ and $N$ very large, we can make the holographic computation reliable in parametrically larger extents of the bulk spacetime. A more detailed analysis of the regime of validity, including finite size effects arising by considering the PFT on a torus, can be found in~\cite{Ganor:2007qh}. In our case, we consider the PFT in a large enough box so that we need not worry about T-duality along the worldvolume. Note however that the volume of a puff is proportional to this box volume; so, we would need a finite but large box with keep things controlled.

\sechead{UV-IR relation and thermodynamics}

In~\cite{Ganor:2007qh}, the finite temperature realization of~(\ref{eq:metric}) was also considered. As usual, it is given by insertions of horizon generating factors in $g_{TT}$ and $g_{ww}$, leading to a black hole with finite temperature 
\begin{equation}
	\mbox{Temp} = \frac{1}{\pi\,{G}^{1/3} \Delta} w_h \ , 
\end{equation}
where $w=w_h$ is the location of the horizon. This helps us identify a UV-IR relation between energy scale $\mu$ in the PFT and extent $w$  in the bulk
\begin{equation}
	\label{eq:uvir} w\equiv {\mu}\, G^{1/3} \Delta\ . 
\end{equation}
Hence, large $w$ corresponds to the UV regime, and small $w$ to the IR.

\sechead{Charges}  \label{sec:charges}

To compute the charges of any D3 brane probe configuration, we will need the Killing vectors of the background geometry. These include generators of the $U(1)\times U(2)$ R-symmetry group of the dual PFT. We have four such generators given by~\cite{FreedBrown:2009py} 
\begin{equation}
	K_0= 
	\partial_\chi \ ; 
\end{equation}
\begin{equation}
	K_1=\cot {\theta} \sin {\varphi}\, 
	\partial_\varphi-\frac{1}{2} \tan (\theta/2) \sin {\varphi}\, 
	\partial_\phi-\cos {\varphi}\, 
	\partial_\theta \ ; 
\end{equation}
\begin{equation}
	K_2=-\cot {\theta} \cos {\varphi}\, 
	\partial_\varphi+\frac{1}{2} \tan (\theta/2) \cos {\varphi}\, 
	\partial_\phi-\sin\varphi\, 
	\partial_\theta \ ; 
\end{equation}
\begin{equation}
	K_3=-\frac{1}{2} 
	\partial_\phi- 
	\partial_\varphi \ ; 
\end{equation}
\begin{equation}
	\label{eq:K4} K_4= 
	\partial_\phi \ . 
\end{equation}
We also have translational symmetries in space $P_i= 
\partial_{X^i}$ and time $E=
\partial_T$. $K_0$ corresponds to the $U(1)$ while $K_1\cdots K_4$ generate the $U(2)$.

Most interestingly, the theory is homogeneous and isotropic despite its non-local attributes. Charge associated with $K_4$ is of particular importance: states that carry this charge are expected to correspond to non-local states~\cite{Ganor:2006ub}\cite{FreedBrown:2009py}, with the scale of non-locality or `puffness' proportional to this charge 
\begin{equation}
	Q_4\propto \frac{\mbox{Volume}}{\Delta^3} 
\end{equation}
where `Volume' refers to the volume of the corresponding non-local state: the non-locality is $SO(3)$ invariant in the $X^1$-$X^2$-$X^3$ subspace; and it is suggested~\cite{Ganor:2006ub} that the corresponding states may be thought of as D3-brane spherical bubbles of finite volume.

\sechead{Holographic screen}

In~\cite{FreedBrown:2009py}, various pieces of evidence were presented suggesting that non-local states may be viewed as `inserted' deep into the holographic bulk, at an extent in the holographic direction $w$ - instead of the boundary at $w\rightarrow\infty$ - given by 
\begin{equation}
	w=w_H=\frac{2^{1/6}}{\cos^{1/3}\psi}\ . 
\end{equation}
The UV-IR map relates this energy scale to the expected extended size of the non-local puffs. We work at the fixed $\psi=0$ plane and this holographic screen would be at $w_H=2^{1/6}$. Note that, at the singularity $\psi=\pi/2$, this screen is pushed deep into the UV. In~\cite{FreedBrown:2009py}, it was also pointed out that this screen appears to split the holographic bulk into two regions, with both sides projecting onto the common boundary at $w_H$: it was proposed that the side with small $w$ holographically encodes dynamics of the non-local states at lengths greater than their puffed-up size; while the large $w$ region encodes the internal dynamics of the puffs. This picture makes sense if we are to view the non-local states as footprints of D3 brane protrusions into the bulk.

\section{Half BPS probes} \label{sec:half_bps_operators}

\subsection{Background Killing spinors} 

\label{sec:killing}

Finding the Killing spinors for our background geometry is a straightforward exercise, albeit slightly more cumbersome than the norm due to the PFT's geometric twist. One finds from~(\ref{eq:killing}) that the Killing spinor is given by 
\begin{equation}
	\label{eq:killingcondition0} \epsilon = \sqrt{w} H^{-1/4}\, M \epsilon_0 
\end{equation}
where 
\begin{equation}
	\label{eq:killingspinor} M\equiv e^{\frac{1}{2} \psi \gamma^{34}} e^{\frac{1}{2} \chi \gamma^{49}} e^{-\frac{1}{2} \phi \gamma^{13}} e^{\frac{1}{2} \phi \gamma^{28}} e^{\frac{1}{4} \theta \gamma^{38}} e^{-\frac{1}{4} \theta \gamma^{12}} e^{-\frac{1}{2} \varphi \gamma^{28}} \ . 
\end{equation}
The numeric indices on the gamma matrices refer to orthonormal tangent space coordinates, {\em i.e.} $\gamma^a = {e^a}_\mu \gamma^\mu$, with the mapping 
\begin{center}
	\begin{tabular}
		{c|c|c|c|c|c|c|c|c|c} $T$ & $\phi$ & $\varphi$ & $w$ & $\psi$ & $X^1$ & $X^2$ & $X^3$ & $\theta$ & $\chi$\\
		\hline 0 & 1 & 2 & 3 & 4 & 5 & 6 & 7 & 8 & 9\\
	\end{tabular}
\end{center}

In particular, the directions parallel to the source D3 branes are $567$ and the PFT twist is in the $1$ direction. $\epsilon_0$ is a constant complex chiral spinor satisfying 
\begin{equation}
	\label{eq:killingcondition} \gamma^{0123456789}\epsilon_0=+\epsilon_0\ \ \ ,\ \ \ (1-i \gamma^{0567}) \epsilon_0 = 0 \ \ \ ,\ \ \ (1 -\gamma^{1238}) \epsilon_0 = 0\ . 
\end{equation}

The first condition projects onto a chiral sector; the second is the usual one for D3 branes without a Melvin twist; the third brings down the supersymmetry from $\mathcal{N}=4$ to $\mathcal{N}=2$ in the dual theory, {\em i.e.} the background has eight supersymmetries. There is no enhanced superconformal symmetry since the non-locality scale breaks conformal invariance. The AdS limit takes $H\rightarrow 1$ in ~(\ref{eq:killingcondition0}) and drops the third condition in~(\ref{eq:killingcondition}).

\subsection{SUSY of probe embeddings} \label{sub:subsection_name}

We want to find BPS embeddings of D3 brane probes with less than eight supersymmetries. Hence, we are to use equation~(\ref{eq:bps}) to find target space embedding functions $x^{\mu}(\sigma)$ using the background Killing spinor given by~(\ref{eq:killingcondition0})-(\ref{eq:killingcondition}) as a starting point. These embeddings may impose additional conditions on $\epsilon_0$ provided that these conditions are compatible with~(\ref{eq:killingcondition}). 

We look for configurations with two translational isometries - potentially planar volume operators extended along $X^2$ and $X^3$. Throughout, we adopt the static gauge. We first choose $\sigma^0\equiv T$, $\sigma^2\equiv X^2$ and $\sigma^3\equiv X^3$. We thus have fixed three of the four worldvolume reparameterization symmetries. The translational isometries imposed on the D3 probe imply\footnote{We use the notation $
\partial_i\equiv
\partial_{\sigma^i}$ throughout.} 
\begin{equation}
	\partial_{2}, 
	\partial_{3} \rightarrow 0 
\end{equation}
for all target space coordinates (except that is $\partial_2 X^2 = 1$, $\partial_3 X^3 = 1$). Furthermore, we restrict to the subspace 
\begin{equation}
	\psi =0 \ \ \ ,\ \ \ \chi = 0 
\end{equation}
for simplicity as well as to avoid the singularity at $\psi=\pi/2$.

We still have one coordinate choice freedom in fixing the static gauge. We can extend the third direction of the probe D3 brane in various target space directions. Certain choices can rule out certain cases, or make determining particular configurations more cumbersome. For the sake of presenting a general treatment, we consider three possible embeddings:

\sechead{Parallel embedding}

With this choice, we stretch the probe parallel to the source D3 brane, choosing the static gauge $\sigma^1=X^1$. This hence cannot lead to volume operators since we would need $X^1=\mbox{constant}$ to realize a co-dimension one defect. The probe may still have protrusions or bents in four transverse directions, described by functions $\phi(\sigma^0,\sigma^1)$, $\varphi(\sigma^0,\sigma^1)$, $w(\sigma^0,\sigma^1)$, and $\theta(\sigma^0,\sigma^1)$.

\sechead{Holographic embedding}

We stretch the probe transverse to the source D3 brane along the holographic direction $w$, choosing the static gauge $\sigma^1=w$. The probe may still have protrusions or bents in $\phi(\sigma^0,\sigma^1)$, $\varphi(\sigma^0,\sigma^1)$, $\theta(\sigma^0,\sigma^1)$, and $X^1(\sigma^0,\sigma^1)$.

\sechead{$\phi$ wrapping}

We wrap the probe along the transverse $\phi$ angle associated with the Melvin twist of the background by choosing the static gauge $\sigma^1=\phi$. The probe may still have protrusions or bents in $\varphi(\sigma^0,\sigma^1)$, $w(\sigma^0,\sigma^1)$, $\theta(\sigma^0,\sigma^1)$, and $X^1(\sigma^0,\sigma^1)$.

\subsection{BPS conditions} \label{sub:bps_conditions}

Given the setup described in the previous section, and after some significant amount of algebra with gamma matrices, one can write the BPS condition arising from~(\ref{eq:bps}) as 
\begin{eqnarray}
	\label{eq:mainbps} \epsilon_0&+&\frac{i \Delta}{\mathcal{L}_{DBI}}\epsilon_0 - \frac{1}{\mathcal{L}_{DBI}} \left(\Delta^{0367} \gamma^{0367} + i \Delta^{03}\gamma^{03} + i \Delta^{58}\gamma^{58} + i \Delta^{08}\gamma^{08} + i \Delta^{15}\gamma^{15} + i \Delta^{01}\gamma^{01} \right. \nonumber \\
	&+&\left.\Delta^{0267}\gamma^{0267} + i \Delta^{02}\gamma^{02} + i \Delta^{0158}\gamma^{0158} + i \Delta^{0135}\gamma^{0135} + i \Delta^{0125}\gamma^{0125} \right) \epsilon_0 = 0\ . 
\end{eqnarray}
In arriving at this expression, we used the explicit form of the background Killing spinor given by~(\ref{eq:killingcondition0})-(\ref{eq:killingcondition}). This structure of the BPS condition is reproduced for all three possible embeddings considered: parallel, holographic, or $\phi$-wrapping. In each case, the $\Delta$'s are complicated functions of the derivatives of the target space coordinates; for the three different wrapping cases, these expressions for the $\Delta$'s differ and are listed in detail in the Appendix for the reader's entertainment. 

Without delving into the details inside the $\Delta$'s, it is straightforward to analyze possible solutions to the BPS condition. Noting that $\epsilon_0$ is a constant complex chiral spinor subject to conditions given by~(\ref{eq:killingcondition}), we can identify only two possible scenarios. Any other possibility conflicts with one or both of the following statements: (1) the corresponding condition $(1+\Gamma)\epsilon_0=0$ for some $\Gamma$ is not a proper projection with $\Gamma^2=1$ and $\mbox{Tr}\, \Gamma=0$; (2) the corresponding condition conflicts with the background Killing conditions~(\ref{eq:killingcondition}) and hence breaks all SUSY's. The remaining two scenarios left are:

\sechead{Scenario I}

We set 
\begin{equation}
	\label{eq:scenarioIa} \epsilon_0+\frac{i \Delta}{\mathcal{L}_{DBI}}\epsilon_0\to 0 
\end{equation}
in ~(\ref{eq:mainbps}). We then need 
\begin{eqnarray}
	\label{eq:scenarioIb} \Delta^{58} = \pm \Delta^{08} &\Rightarrow & (1\mp \gamma^{05})\epsilon_0 = 0 \nonumber \\
	\Delta^{15} = \pm \Delta^{01} &\Rightarrow & (1\pm \gamma^{05})\epsilon_0 = 0 \nonumber \\
	\Delta^{0367} = \pm \Delta^{03} &\Rightarrow & (1\mp i\,\gamma^{67})\epsilon_0 = 0 \nonumber \\
	\Delta^{0267} = \pm \Delta^{02} &\Rightarrow & (1\mp i\,\gamma^{67})\epsilon_0 = 0 
\end{eqnarray}
For this to work, we would then need all the remaining $\Delta$'s to vanish 
\begin{equation}
	\label{eq:scenarioIc} \Delta^{0158}=\Delta^{0135}=\Delta^{0125} = 0 
\end{equation}
to have $1/2$ BPS configurations. Note that the two conditions on $\epsilon_0$ appearing in~(\ref{eq:scenarioIb}) are related: 
\begin{equation}
	(1\mp i \gamma^{67})\epsilon_0 = 0 \Rightarrow (1\mp\gamma^{05})\epsilon_0 = 0 
\end{equation}
because we always have $(1- i \gamma^{0567})\epsilon_0 = 0$ from ~(\ref{eq:killingcondition}). Hence, this scenario gives probes with four supersymmetries. The structure of the BPS condition suggests we may be dealing with waves in the $5$ (or $X^1$) direction.

Equations~(\ref{eq:scenarioIa})-(\ref{eq:scenarioIc}) lead to a system of first order differential equations for the embedding functions after using the explicit forms of the $\Delta$'s listed in the Appendix. The task then becomes to solve these differential equations and check that they lead to consistent and real solutions for the embedding.

\sechead{Scenario II}

We set 
\begin{equation}
	\label{eq:scenarioIIa} \Delta\to 0 
\end{equation}
in ~(\ref{eq:mainbps}) with $\mathcal{L}_{DBI}\neq 0$. We then have three possibilities: 
\begin{eqnarray}
	\label{eq:scenarioIIb} \mbox{Scenario IIa: }\ \Delta^{0135}=\pm \mathcal{L}_{DBI}\ \ \ ,\ \ \ \Delta^{0125}=\Delta^{0158}=0 \Rightarrow (1\mp i \gamma^{0135})\epsilon_0=0 \nonumber \\
	\mbox{Scenario IIb: }\ \Delta^{0125}=\pm \mathcal{L}_{DBI}\ \ \ ,\ \ \ \Delta^{0135}=\Delta^{0158}=0 \Rightarrow (1\mp i \gamma^{0125})\epsilon_0=0 \nonumber \\
	\mbox{Scenario IIc: }\ \Delta^{0158}=\pm \mathcal{L}_{DBI}\ \ \ ,\ \ \ \Delta^{0125}=\Delta^{0135}=0 \Rightarrow (1\mp i \gamma^{0158})\epsilon_0=0 
\end{eqnarray}
with 
\begin{equation}
	\label{eq:scenarioIIc} \Delta^{58} = \Delta^{08} = \Delta^{15} = \Delta^{01} = \Delta^{0367} = \Delta^{03} = \Delta^{0267} = \Delta^{02} = 0\ . 
\end{equation}
These are three distinct cases. Again, each gives a system of first order differential equations for the embedding functions that we would need to solve.

\sechead{Summary}

In total, we then have four possibilities for $1/2$ BPS configurations to check for: 
\begin{eqnarray}
	\label{eq:scenarios} &&\mbox{Scenario I : }(1\pm i \gamma^{67})\epsilon_0 = 0 \Rightarrow (1\pm\gamma^{05})\epsilon_0 = 0 \nonumber \\
	&&\mbox{Scenario IIa : }(1\pm i \gamma^{0135})\epsilon_0=0 \nonumber \\
	&&\mbox{Scenario IIb : }(1\pm i \gamma^{0125})\epsilon_0=0 \nonumber \\
	&&\mbox{Scenario IIc : }(1\pm i \gamma^{0158})\epsilon_0=0 
\end{eqnarray}
In each case, we have eight partial differential equations for four functions of two variables. At this stage, we are {\em not} guaranteed solutions for any of these cases. We still need to make sure that the differential equations are consistent with each other and lead to real solutions.

\sechead{$1/4$ BPS configurations}

It is easy to see that the two scenarios I and II listed above are also compatible with each other. Hence, we can write three possible conditions for $1/4$ BPS configurations by imposing 
\begin{eqnarray}
	&&\Delta = 0 \nonumber \\
	&&\Delta^{58} = \pm \Delta^{08} \Rightarrow (1\mp \gamma^{05})\epsilon_0 = 0 \nonumber \\
	&&\Delta^{15} = \pm \Delta^{01} \Rightarrow (1\pm \gamma^{05})\epsilon_0 = 0 \nonumber \\
	&&\Delta^{0367} = \pm \Delta^{03} \Rightarrow (1\mp i\,\gamma^{67})\epsilon_0 = 0 \nonumber \\
	&&\Delta^{0267} = \pm \Delta^{02} \Rightarrow (1\mp i\,\gamma^{67})\epsilon_0 = 0 \nonumber \\
	&&\mbox{and}\left \{ 
	\begin{array}{l}
		\mbox{ }\,\,\,\,\,\Delta^{0135}=\pm \mathcal{L}_{DBI}\ \ \ ,\ \ \ \Delta^{0125}=\Delta^{0158}=0 \Rightarrow (1\pm i \gamma^{0135})\epsilon_0=0 \\
		\mbox{or }\Delta^{0125}=\pm \mathcal{L}_{DBI}\ \ \ ,\ \ \ \Delta^{0135}=\Delta^{0158}=0 \Rightarrow (1\pm i \gamma^{0125})\epsilon_0=0 \\
		\mbox{or }\Delta^{0158}=\pm \mathcal{L}_{DBI}\ \ \ ,\ \ \ \Delta^{0125}=\Delta^{0135}=0 \Rightarrow (1\pm i \gamma^{0158})\epsilon_0=0 
	\end{array}
	\right. 
\end{eqnarray}
However, we will not consider such cases in this work.

\sechead{Static configurations}

Before delving into the details of the BPS configurations listed above, let us take a step back and consider in some generality {\em static} configurations. We impose $
\partial_0 = 0$ on all coordinates (except that is $
\partial_0 T= 1$); and relax the conditions $\psi=0$ and $\chi=0$ as well. Furthermore, we relax the fourth gauge fixing condition that restricts to parallel, holographic, or $\phi$ embeddings: we want as general a treatment as possible within the static scenario. Hence, we are considering static probe configurations with two translational isometries along the worldvolume of the source branes; but otherwise of arbitrary shape. This setup leads to the BPS condition 
\begin{eqnarray}
	\epsilon&-& \frac{1}{H\,\mathcal{L}_{DBI}}\left( \frac{i 
	\partial_1R}{H}\gamma^{567} -{i w 
	\partial_1\chi \sin \psi}\gamma ^{679}-\frac{i w }{2}{
	\partial_1\theta \cos \psi}\gamma^{678}-{i w 
	\partial_1\psi}\gamma^{467} \right. \nonumber \\
	&-&\left.\frac{i w}{2}{
	\partial_1\varphi \sin \theta \cos \psi}\gamma^{267} -\frac{i w}{2}{\cos \psi \left(
	\partial_1\varphi (\cos\theta-1)+2\, 
	\partial_1\phi \right)}\gamma^{167} -{i\, {
	\partial_1 w}\gamma}^{367} \right)\epsilon = 0\ . 
\end{eqnarray}
Using the form of the Killing spinor given in~(\ref{eq:killingcondition0})-(\ref{eq:killingcondition}), one can check - after some amount of unpleasant algebra - that there are no non-trivial embeddings that solve this BPS condition because one cannot find a proper projection operator compatible with the background Killing spinor. Hence, {\em there are no non-trivial BPS static configurations of probe D3 branes in this background.} This is probably due to the fact that the background is not static, albeit stationary: the spacetime is spinning along $\phi$ and $\varphi$ and, to keep an embedding `in place', one needs to have it move as well.

\section{Results} \label{sec:section_name}

\subsection{Setup} \label{subsec:action_and_charges}

In this section, we present the details of 1/2 BPS configurations for D3 brane probes in the holographic PFT background. The list is exhaustive for all configurations satisfying the following conditions: (1) all embeddings have two translational isometries along the PFT branes; (2) all configurations lie in the $\psi=\chi=0$ plane. We also have ruled out any solutions that are static and have two translational isometries.

We present only the cases that lead to distinct physical embeddings. For each case at hand, we show the first order BPS differential equations and general solutions to the equations; and if the configuration is a candidate volume operator of the PFT, we also compute the action, the energy, and R-charge. 

In computing the action, we need to add to it a boundary term to obtain a good variational principle. This boundary term takes the form (see for example~\cite{Drukker:2008wr})
\begin{equation}
	\mathcal{L}\rightarrow \mathcal{L}-\frac{
	\partial \mathcal{L}}{
	\partial {x^\mu}'} {x^{\mu}}'\ . 
\end{equation}
The action computed from this modified Lagrangian is expected to be related to the vacuum expectation value of a dual operator $\mathcal{O}$ 
\begin{equation}
	\langle \mathcal{O} \rangle \propto e^{i \mathcal{S}}\ . 
\end{equation}
The energy is then computed using the Noether method and the Killing vector $
\partial_T$ 
\begin{equation}
	\label{eq:energy} E = \int d \sigma^1 \frac{
	\partial \mathcal{L}}{
	\partial (
	\partial_0 T)} 
\end{equation}
where $E$ is energy per unit area (since the $X^2$ and $X^3$ directions of the probe extend to infinity). Note that this expression is used {\em before} fixing the static gauge. This energy would be related to the mass dimension of the corresponding operator. The R-charge of interest is the one associated with $K_4$ in~(\ref{eq:K4}), {\em i.e.} momentum along the twist direction $\phi$; the charge per unit area is then given by 
\begin{equation}
	\label{eq:charge} Q_4 = \int d \sigma \frac{
	\partial \mathcal{L}}{
	\partial (
	\partial_0 x^\mu)} K_4^\mu = \int d \sigma \frac{
	\partial \mathcal{L}}{
	\partial (
	\partial_0 \phi)}\ . 
\end{equation}

\subsection{Scenario I: A warm-up exercise} \label{sub:scenarioI}

For this case, we need to impose~(\ref{eq:scenarioIa})-(\ref{eq:scenarioIc}) and we have one BPS condition at hand, scenario I of ~(\ref{eq:scenarios}). We are solving for the four embedding functions $w(\sigma^0,\sigma^1)$, $\theta(\sigma^0,\sigma^1)$, $\phi(\sigma^0,\sigma^1)$, and $\varphi(\sigma^0,\sigma^1)$. We choose a parallel embedding for convenience, with $\sigma^0=T$ and $\sigma^1=X^1$. After using the explicit form of the $\Delta$'s listed in the appendix, we find simply 
\begin{equation}
	(
	\partial_0 \mp 
	\partial_1) w= 0,\ \ (
	\partial_0 \mp 
	\partial_1) \theta= 0 ,\ \ (
	\partial_0 \mp 
	\partial_1) \phi = 0 ,\ \ (
	\partial_0 \mp 
	\partial_1) \varphi = 0 
\end{equation}
with the BPS condition 
\begin{equation}
	(1\mp\gamma^{05})\epsilon_0 = 0\ . 
\end{equation}
The configuration then describes waves with four arbitrary profile functions $w(\sigma^\pm)$, $\theta(\sigma^\pm)$, $\varphi(\sigma^\pm)$, $\phi(\sigma^\pm)$ - with all functions depending on $\sigma^\pm=\sigma^0\pm\sigma^1=T\pm X^1$ only. These are also BPS configurations for the $AdS_5\times S^5$ background. They are not surface/volume operators; they correspond to adding a wave on a parallel probe brane breaking an additional $1/2$ of the supersymmetries as is commonly known. This is not the focus of this work and we will instead move onto configurations that have attributes of volume operators in the PFT. 

\subsection{Scenario II: Volume operators} \label{sub:scenarioII}

We are now considering the conditions~(\ref{eq:scenarioIIa})-(\ref{eq:scenarioIIc}). We have three cases to consider, scenarios IIa, IIb, and IIc of~(\ref{eq:scenarios}).

\subsubsection{Scenario IIa: $(1+ i\,\gamma^{0513})\epsilon_0 = 0$} \label{subsub:scenarioIIa}

\begin{figure}
	\begin{center}
		\includegraphics[width=6.5in]{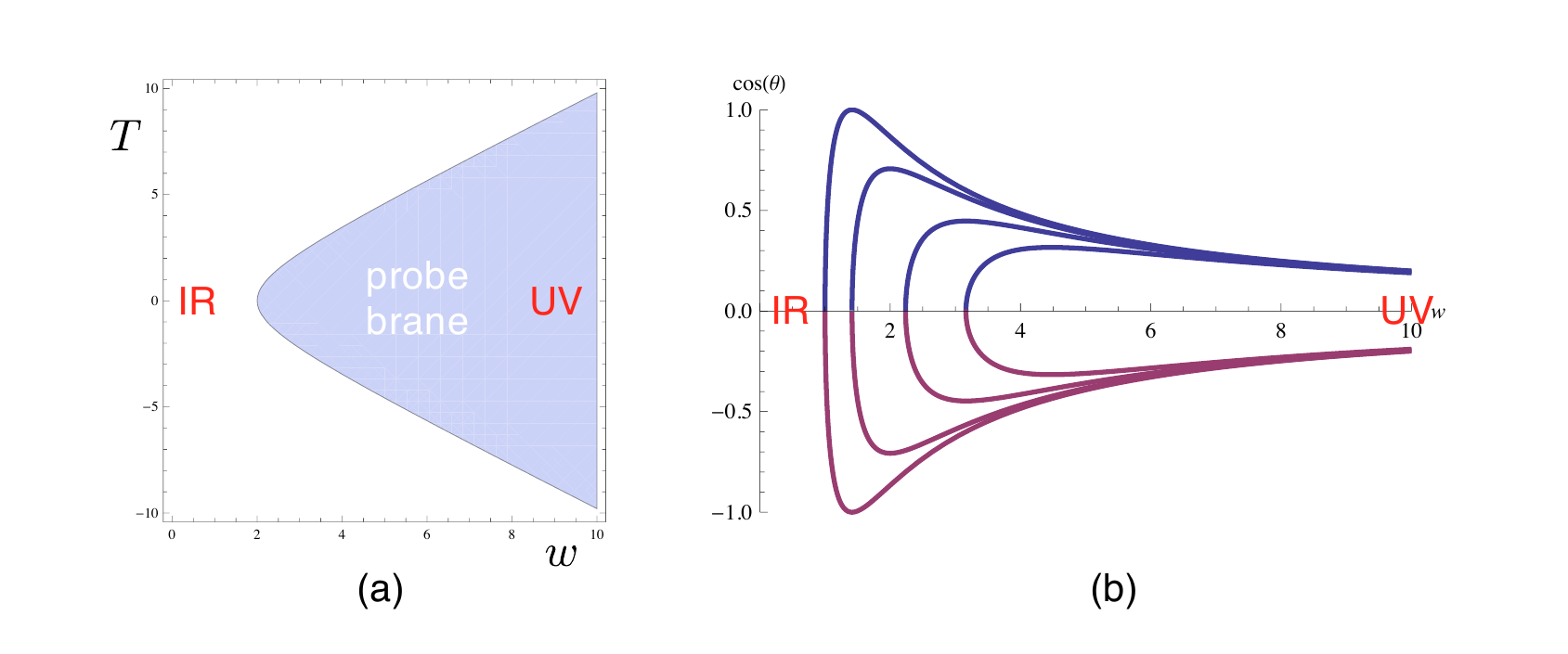} 
	\end{center}
	\caption{\em (a) A profile of the D3 probe in the $T$ vs $w$ plane; the shaded area is the worldvolume of the D3 branes; (b) The probe as seen from the $\cos\theta$ versus $w$ plane for $C_3=1/2$ and various time snapshots $T=0,1,2,3$.} \label{fig:fig21-profile1} 
\end{figure}

\begin{figure}
	\begin{center}
		\includegraphics[width=6.5in]{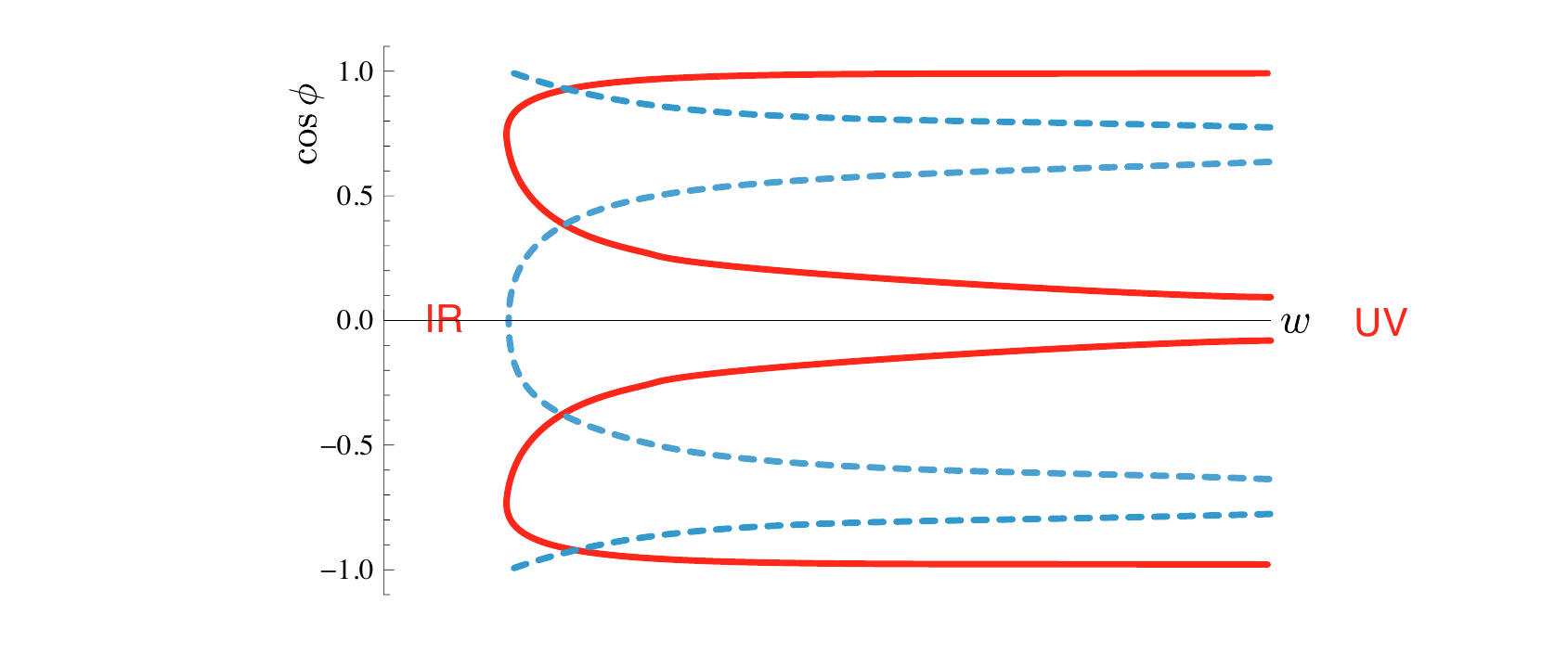} 
	\end{center}
	\caption{\em The branches of $\cos\phi$ versus $w$ for the clover defect of scenario IIa. There are two separate configurations depicted as solid and dashed lines. They are related by a rotation by $\pi/4$ in $\phi$.} \label{fig:fig21-profileangles1} 
\end{figure}

For this case, we consider a holographic embedding $\sigma^0=T$ and $\sigma^1=w$ for convenience. Equations~(\ref{eq:scenarioIIa})-(\ref{eq:scenarioIIc}) lead to (after using the explicit expressions for the $\Delta$ from the Appendix) 
\begin{eqnarray}
	&&
	\partial_0 X^1 = 0\ \ \ ,\ \ \ 
	\partial_1 X^1= 0,\nonumber \\
	&&
	\partial_0 \phi = \frac{1}{w^2}+\sin ^2 \frac{\theta}{2}\,\, 
	\partial_0 \varphi \ \ \ ,\ \ \ 
	\partial_1 \phi = \sin ^2 \frac{\theta}{2}\,\, 
	\partial_1 \varphi \nonumber \\
	&&
	\partial_0 \varphi = \frac{w^4 (
	\partial_0 \theta)^2+2\, w\, 
	\partial_1 \theta \cot \theta+4}{w^2 \left(2 \cos \theta-w\, 
	\partial_1 \theta \sin \theta\right)}\ \ \ ,\ \ \ 
	\partial_1 \varphi = -w \csc \theta\,\,
	\partial_0 \theta \ . 
\end{eqnarray}
Note in particular that we have $X^1=\mbox{constant}$: this is an embedding extending in the $X^2$ and $X^3$ directions at fixed $X^1$ in the dual PFT, {\em i.e.} potentially a planar defect or volume operator - if it extends to the UV boundary. There is only one BPS condition possible 
\begin{equation}
	(1+i \gamma^{0135})\epsilon_0 = 0\ . 
\end{equation}
We have checked that this is a new case that does not exist in the case of $AdS_5\times S^5$. We then need to double check whether the system leads to a real physical embedding of the D3 brane probe. We start by requiring $
\partial_0
\partial_1=
\partial_1
\partial_0$ on all the functions. This leads to a separation of variables in $\theta$ 
\begin{eqnarray}
	&&w^4 (
	\partial_0\theta ){}^2 \cos \theta+w^2 (
	\partial_1\theta ){}^2 \cos \theta+4 w \sin \theta\,\, 
	\partial_1 \theta -4 \cos \theta=0, \nonumber \\
	&&-w^2 
	\partial_{1}^2 \theta-2 w^2 (
	\partial_1\theta )^2 \cot (2 \theta )-7\, w\, 
	\partial_1 \theta+4 \cot \theta=0 
\end{eqnarray}
Albeit non-linear in $\theta(w,T)$, these equations are exactly solvable. We find 
\begin{eqnarray}
	\cos \theta(T,w) = \pm\frac{\sqrt{C_1(T)} \sqrt{16 w^2 C_2(T)-1}}{\sqrt{2} w^2} 
\end{eqnarray}
where the two functions of time $C_1(T)$ and $C_2(T)$ must then satisfy 
\begin{eqnarray}
	&&C_2(T) C_1(T)'+C_1(T) C_2(T)'=0, \nonumber \\
	&&C_1(T)'=\pm \sqrt{32\, C_1(T)} \sqrt{1-32 C_1(T) C_2(T)^2}\ . 
\end{eqnarray}
Once again, these are solvable in closed form, and we find for $\theta$ 
\begin{equation}
	\cos \theta(T,w) = \pm\frac{4\,C_3}{w^2}\sqrt{\frac{w^2}{2\,C_3}-\frac{T^2}{4\,C_3^2}-1} 
\end{equation}
with a constant of integration $C_3$. Another constant that simply shifts time has been set to zero. Having $\theta(T,w)$, we can now find closed form solutions for $\varphi(T,W)$ and $\phi(T,w)$ as well: 
\begin{eqnarray}\label{eq:scenIvarphi}
	&&\varphi(T,w) = C_5\mp\frac{1}{2} \tan ^{-1}\left[\frac{2 T \left(4 C_3^2-C_3 w^2+T^2\right) \sqrt{-4 C_3^2+2 C_3 w^2-T^2}}{\left(4 C_3^2-C_3 w^2+2 T^2\right)^2-2 T^2 \left(2 C_3^2+T^2\right)}\right] 
\end{eqnarray}
and 
\begin{eqnarray}\label{eq:scenIphi}
	\phi(T,w) &=& C_6+\frac{1}{4} \tan ^{-1}\left[\frac{4 T \left(w^2-4 C_3\right)}{4 T^2-\left(w^2-4 C_3\right)^2}\right] \nonumber \\
	&\mp &\frac{1}{4} \tan ^{-1}\left[\frac{2 T \left(4 C_3^2-C_3 w^2+T^2\right) \sqrt{-4 C_3^2+2 C_3 w^2-T^2}}{\left(4 C_3^2-C_3 w^2+2 T^2\right)^2-2 T^2 \left(2 C_3^2+T^2\right)}\right] 
\end{eqnarray}
where $C_5$ and $C_6$ are constants of integration. In total, we have four constants of integration, $C_3$, $C_5$, $C_6$, and a time shift parameter we have set to zero. Making sure that $|\cos\theta(T,w)| < 1$, and that these expressions lead to real solutions, we find a single unifying condition 
\begin{equation}
	\label{eq:capIIa} \frac{w^2}{2\, C_3}-\frac{T^2}{4\, C_3^2}>1\ . 
\end{equation}
Hence, $w$ is bounded from below, in the IR, by 
\begin{equation}
	w_{min}= \sqrt{2\,C_3} 
\end{equation}
with $C_3>0$. Figure~\ref{fig:fig21-profile1} depicts sections of the profile. 
The probe extends to the UV boundary $w\rightarrow \infty$ of the bulk space and lands on it as a plane extended in the $X^2$ and $X^3$ directions, at fixed $X^1$. However, in the angular directions $\theta$, $\phi$, and $\varphi$, the configurations expands and is a folding D3 brane, smoothly capping off in the IR at a fixed value in $w$ given by~(\ref{eq:capIIa}). The cap in the IR hence moves with time along the trajectory shown in Figure~\ref{fig:fig21-profile1}(a). And we can see the cap in the $\theta$-$w$ plane in Figure~\ref{fig:fig21-profile1}(b) at various snapshots in time. The $\varphi$ and $\phi$ directions have multiple branches due to the inverse tangent functions in equations~(\ref{eq:scenIvarphi}) and~(\ref{eq:scenIphi}): for $\varphi$, these branches are reached by adding integer multiples of $\pi/2$ to it (since the arctan is undetermined up to $\pm\pi$); and similarly the branches of $\phi$ are reached by adding integer multiples of $\pi/4$ to it. We also note that near the UV boundary $w\rightarrow \infty$, we have 
\begin{equation}
	\theta \rightarrow \frac{\pi}{2}\ \ \ ,\ \ \ \cos\varphi \rightarrow 0, \pm 1 \ \ \ ,\ \ \ \cos\phi \rightarrow 0,\pm 1\ . 
\end{equation}
where we have set $C_5=C_6=0$ for simplicity. $\theta$ hence lands in the UV at the equator of the $\theta$-$\varphi$ 2-sphere. It is easy to see that the $\varphi$ profile looks qualitatively similar to that of $\theta$ shown in Figure~\ref{fig:fig21-profile1}(b). In particular, in the UV the probe asymptotes to any of $\varphi\rightarrow 0,\pi/2,\pi$, capping off in the IR at one of $\varphi=\pi,\pi/2,0$ respectively. Along each branch, as we move from the UV to the IR and back to the UV, $\theta$ moves from the equator to the North pole to the South pole and back to the equator; and $\varphi$ makes a full circular trip with four possible endpoints in the UV, $0$, $\pi/2$, $\pi$, and $3\pi/2$. The more interesting profile is in the $\phi$ direction. Figure~\ref{fig:fig21-profileangles1} (the solid line)
shows a plot of the various branches of $\phi$. Note in particular that that the probe straddles the Melvin angle $\phi$ in increments of $\pi/2$ only. We can then summarize all these observations with Figure~\ref{fig:main}(a): the D3 probe has a clover profile in the Melvin angle direction, with four branches. Looking back at the metric~(\ref{eq:metric}), we see that the size of the $\phi$ circle goes as
\begin{equation}
	\frac{\alpha' \sqrt{G}}{w^3} d\phi^2\rightarrow 0
\end{equation}
as $w\rightarrow\infty$ in the UV. To determine whether all four branches of the clover converging on the defect land at the same point in the full {\em ten} dimensional space requires considering the T-dual configuration. Since the probe is not wrapped along the T-duality circle $\phi$, it becomes a D4 brane. And since the discontinuity of $\pi/2$ in the UV is in the $\phi$ direction, the T-dual picture will see a stack of {\em eight} D4 branes landing on the now co-dimension one defect (which is on the worldvolume of source D4 branes as well). The branes however can still be separated in the $\varphi$ direction. We may expect that we are dealing with volume operators with maximally $U(8)$, $U(4)\times U(4)$, $U(4)\times U(2)\times U(2)$, or $U(2)\times U(2)\times U(2)\times U(2)$ non-Abelian degrees of freedom - depending on which of the four branches of $\varphi$ gets picked up along the four leaves of the `clover'\footnote{Note that the invariant length along $\varphi$ is large in the UV as can be seen from the metric~(\ref{eq:metric}).}. We will come back to this issue in the Discussion section. We refer to the this BPS profile of the D3 brane probe as the `clover configuration'. 

Note also that the IR cap in the bulk never reaches the boundary: for large $T\rightarrow \pm \infty$, we have from~(\ref{eq:capIIa}) 
\begin{equation}
	w_{cap} \rightarrow \frac{|T|}{\sqrt{2\,C_3}}\ . 
\end{equation}
It takes infinite time for the cap to reach the UV boundary. The solution has four constant parameters: three trivial ones involving translations in $T$, $\phi$, or $\varphi$; plus $C_3$ which tunes the distance in the bulk the probe extends to before folding back onto the defect.

The action - before subtraction of boundary terms and evaluated at this solution - is 
\begin{equation}
	\mathcal{S} = -i \int dT\,d^2X dw \frac{w^4 (
	\partial_0\theta ){}^2+w^2 (
	\partial_1\theta)^2+4}{2\, w \left(w\, 
	\partial_1 \theta \sin \theta -2 \cos (\theta )\right)}\ . 
\end{equation}
After subtraction, the action does not appear to vanish at first 
\begin{equation}
	\mathcal{S} = -i \int dT\,d^2X dw\frac{\left(w^4 (
	\partial_0\theta )^2 \sin ^2 \theta +4\right)}{2\, w \left(w\, 
	\partial_1 \theta \sin \theta -2 \cos \theta \right)} 
\end{equation}
Numerical integration in $w$ from the cap to the UV boundary however shows that this expression does not vanish
\begin{equation}
	\mathcal{S} = \frac{4\,i}{\sqrt{2 C_3}} T_0 A w_{uv}
\end{equation}
where $A$ is the regularized area of the defect in the $X^2$ and $X^3$ directions, the time integral is between $-T_0$ and $T_0$ with $T_0$ large, and $w_{uv}$ is the cutoff edge of the probe at the UV boundary. Hence, the corresponding volume operator's vev is 
\begin{equation}
	\langle \mathcal{O} \rangle \simeq e^{i\mathcal{S}} \simeq e^{-\frac{4}{\sqrt{2 C_3}} T_0 A w_{uv}} = e^{-2 \frac{{V}}{G\Delta^3} \frac{w_{uv}}{w_{ir}}}
\end{equation}
where we have written the result in terms of the physical PFT parameters, with ${V}$ being the `volume' $2\, t_0 A$, and $w_{ir}=\sqrt{2 C_3}$ being the maximum extent of the probe in the deep IR. This has a suggestive form reminiscent of Wilson operators: a volume {\em per puff volume $G\Delta^3$}, and a regularization factor in the holographic direction.

We next evaluate the energy (without subtracting any boundary terms) using the Noether method and the Killing vector $
\partial_T$ as in~(\ref{eq:energy}). We find that the energy vanishes as well, 
\begin{equation}
	E=0\ .
\end{equation}
We expect a simple relation between this energy, the mass dimension of the operator, and the R-charge - given that the configuration is BPS. Since the energy vanishes, we then compute the R-charge.

Using~(\ref{eq:charge}), we also compute the R-charge density $Q_4$ and find 
\begin{equation}
	Q_4= -\int dw \frac{\left(w^6+1\right) \sec ^2({\theta }/{2})}{2\,w\, \mathcal{L}_{DBI}} \left( w^2 (
	\partial _0\theta )^2 \left(2 \, w^2+\cos \theta-1\right)+(\cos \theta+1) \left(w^2 (
	\partial_1\theta )^2+4\right) \right)\ . 
\end{equation}
It is difficult to obtain a closed form expression for this charge; however, we can determine numerically that it is indeed non-zero. We conclude $Q_4\neq 0$: the defect carries R-charge of the type associated with non-local states in the PFT. Given that the state is BPS and the energy vanishes, the mass dimension of the corresponding operator is probably determined by the R-charge. However, we cannot obtain an expression for the mass dimension of the corresponding operator - if one can be defined at all given the non-local character of the insertion.

\subsubsection{Scenarios IIb and IIc: $(1- i\,\gamma^{0512})\epsilon_0 = 0$ and $(1+ i\,\gamma^{0518})\epsilon_0 = 0$} \label{subsub:scenarioIIbc}

For convenience, we now switch to a $\phi$-wrapping gauge: $\sigma^0=T$ and $\sigma^1=\phi$. We start with case IIb in~(\ref{eq:scenarios}). After using the detailed forms of the $\Delta$'s from the Appendix in~(\ref{eq:scenarioIIa})-(\ref{eq:scenarioIIc}), we arrive at the system of differential equations 
\begin{eqnarray}
	\label{eq:equationsIIb} &&
	\partial_0 X^1 = 0\ \ \ ,\ \ \ 
	\partial_1 X^1 = 0, \nonumber \\
	&&
	\partial_0 w = -\frac{\cot (\phi -\varphi )}{w}\ \ \ ,\ \ \ 
	\partial_1 w = w \cos ^2 \frac{\theta}{2} \csc \phi \sin \varphi \csc (\phi -\varphi ), \nonumber \\
	&&
	\partial_0 \varphi = \frac{2}{w^2 (\cos \theta-1)}\ \ \ ,\ \ \ 
	\partial_1 \varphi = -\frac{2}{\cos \theta-1} \nonumber \\
	&&
	\partial_0 \theta = -\frac{2 \cot (\theta/2) \cot (\phi -\varphi )}{w^2}, \nonumber \\
	&&
	\partial_1 \theta = 2 \cot \frac{\theta}{2} \cot (\phi -\varphi )-2 \sin \frac{\theta }{2} \cos \frac{\theta}{2} \csc \phi \sin \varphi \csc (\phi -\varphi )\ . 
\end{eqnarray}
Once again, we note that we have $X^1=\mbox{constant}$: this is an embedding extending in the $X^2$ and $X^3$ directions at fixed $X^1$ in the dual PFT, {\em i.e.} potentially a planar volume operator. The BPS condition is 
\begin{equation}
	(1-i \gamma^{0125})\epsilon_0 = 0\ . 
\end{equation}
We have checked that this is a new case that does not exist in the case of $AdS_5\times S^5$. It is easy to verify that the system of equations is a consistent one with $
\partial_0 
\partial_1 = 
\partial_1 
\partial_0$ acting on any of $w$, $\varphi$, and $\theta$ without any further constraints. Shuffling around these equations, we can write 
\begin{equation}
	\frac{
	\partial_0 w }{w}-\frac{1}{2} 
	\partial_0 \theta \tan \frac{\theta}{2}= 0 
\end{equation}
which leads to 
\begin{equation}
	w= f(\phi ) \sec (\theta/2) 
\end{equation}
for some unknown function $f(\phi)$. Going back to~(\ref{eq:equationsIIb}), we can rewrite one of the equations as 
\begin{equation}
	w^2 
	\partial_0 \theta+
	\partial_1 \theta +2\, \frac{
	\partial_1 w }{w} \tan \frac{\theta}{2}=0 
\end{equation}
which can be rearranged as 
\begin{equation}
	\frac{f(\phi )'}{f(\phi )^3}+\frac{
	\partial_1\left(\log \left(\tan \left({\theta }/{2}\right)\right)\right)}{f(\phi )^2}+
	\partial_0\left[\log \tan \frac{\theta }{2}\right]= 0\ . 
\end{equation}
This allows us to solve for $\theta$ in closed form 
\begin{equation}
	\cos \theta= \frac{2 f(\phi )^2}{f(\phi )^2+g\left(\int^{\phi } f(x)^2 \, dx-T\right){}^2}-1 
\end{equation}
in terms of another arbitrary function $g(\cdot)$. Let's define 
\begin{equation}
	\Phi \equiv \int^{\phi } f(x)^2 \, dx-T 
\end{equation}
as a shorthand. Going back to~(\ref{eq:equationsIIb}), we have 
\begin{equation}
	\partial_1 \varphi = \frac{f(\phi )^2}{g\left(\Phi\right){}^2}+1 
\end{equation}
or 
\begin{equation}
	\varphi = \int^{\phi } \frac{f(y)^2}{g\left(\Phi\right){}^2} \, dy+h(T)+\phi 
\end{equation}
for yet another undetermined function $h(T)$. We then write $
\partial_0\varphi$ from~(\ref{eq:equationsIIb}), which now looks like 
\begin{equation}
	h(T)'= -\frac{1}{g\left(\Phi\right){}^2}-\int^{\phi } \frac{2 f(y)^2 g\left(\Phi\right)'}{g\left(\Phi\right){}^3} \, dy 
\end{equation}
which implies 
\begin{equation}
	h(T)'= 0 \Rightarrow h(T)=C_1\ . 
\end{equation}
This is because the left hand side is only a function of $T$. We now look at $
\partial_0 w$ in~(\ref{eq:equationsIIb}), and we get 
\begin{equation}
	-g\left(\Phi\right) g\left(\Phi\right)'=\cot \left[ \int^{\phi } \frac{f(y)^2}{g\left(\Phi\right){}^2} \, dy + C_1 \right]\ . 
\end{equation}
Taking the derivative of this with respect to $\phi$ leads to the simple equation 
\begin{equation}
	f(\phi )^2 \left(1-g\left(\Phi\right){}^3 g''\left(\Phi\right)\right)=0\ . 
\end{equation}
We then have two possibilities 
\begin{equation}
	f(\phi ) = 0 \ \ \ \mbox{or}\ \ \ 1-g\left(\Phi\right){}^3 g''\left(\Phi\right)=0\ . 
\end{equation}
In either case, we have one left over unknown function - either $f$ or $g$ - to solve for, and exactly one equation not yet used from~(\ref{eq:equationsIIb}): 
\begin{equation}
	\label{eq:remainingII} 
	\partial_1 w=w \csc \phi \,\sin \varphi \, \csc (\phi -\varphi )\,\cos ^2\frac{\theta }{2}\ . 
\end{equation}
For the first possibility, $f(\phi ) = 0$, we have the embedding 
\begin{equation}\label{eq:c1}
	w^2 = g(-T)^2\ \ \ ,\ \ \ \cos\theta = -1\ \ \ ,\ \ \ \varphi = \phi+C_1\ . 
\end{equation}
We see that~(\ref{eq:remainingII}) is satisfied and the remaining function $g(-T)$ remains arbitrary. The probe D3 brane extends in the $X^2$ and $X^3$ directions while wrapping the $\varphi$ cycle at fixed $\theta=\pi$ (at the south pole). It has fixed $X_1$ coordinate but - in the holographic direction $w$ - it is also at a fixed point that can change in time. This means this possibility does not lead to a volume operator that lands on a planar defect at the boundary unless $w=g\rightarrow \infty$. 

\begin{figure}
	\begin{center}
		\includegraphics[width=6in]{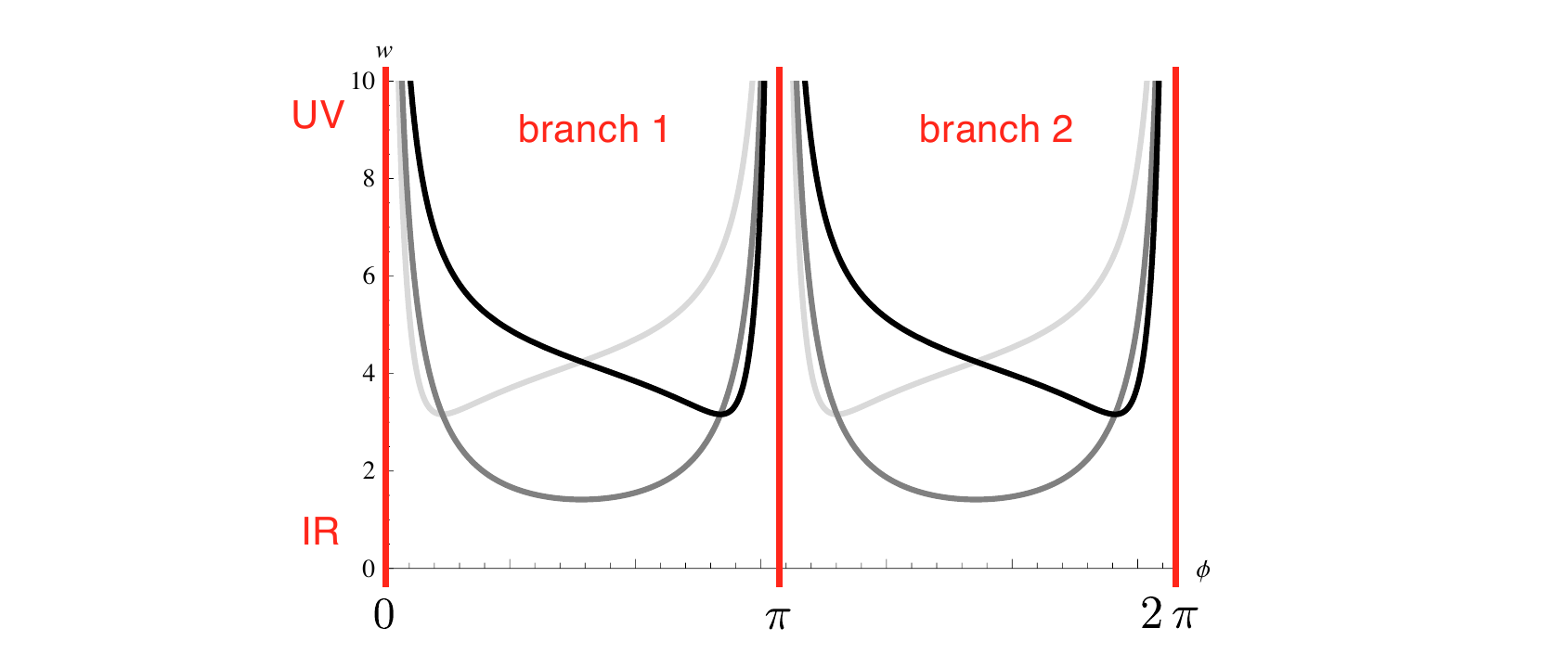} 
	\end{center}
	\caption{\em The probe D3 brane in the $w$ versus $\phi$ plane with $C_2=1$, $C_3=0$, and $C_4=1$, and for time snapshots of $T=-4,0,4$ from light gray to black.} \label{fig:fig22-profile} 
\end{figure}

\begin{figure}
	\begin{center}
		\includegraphics[width=6.5in]{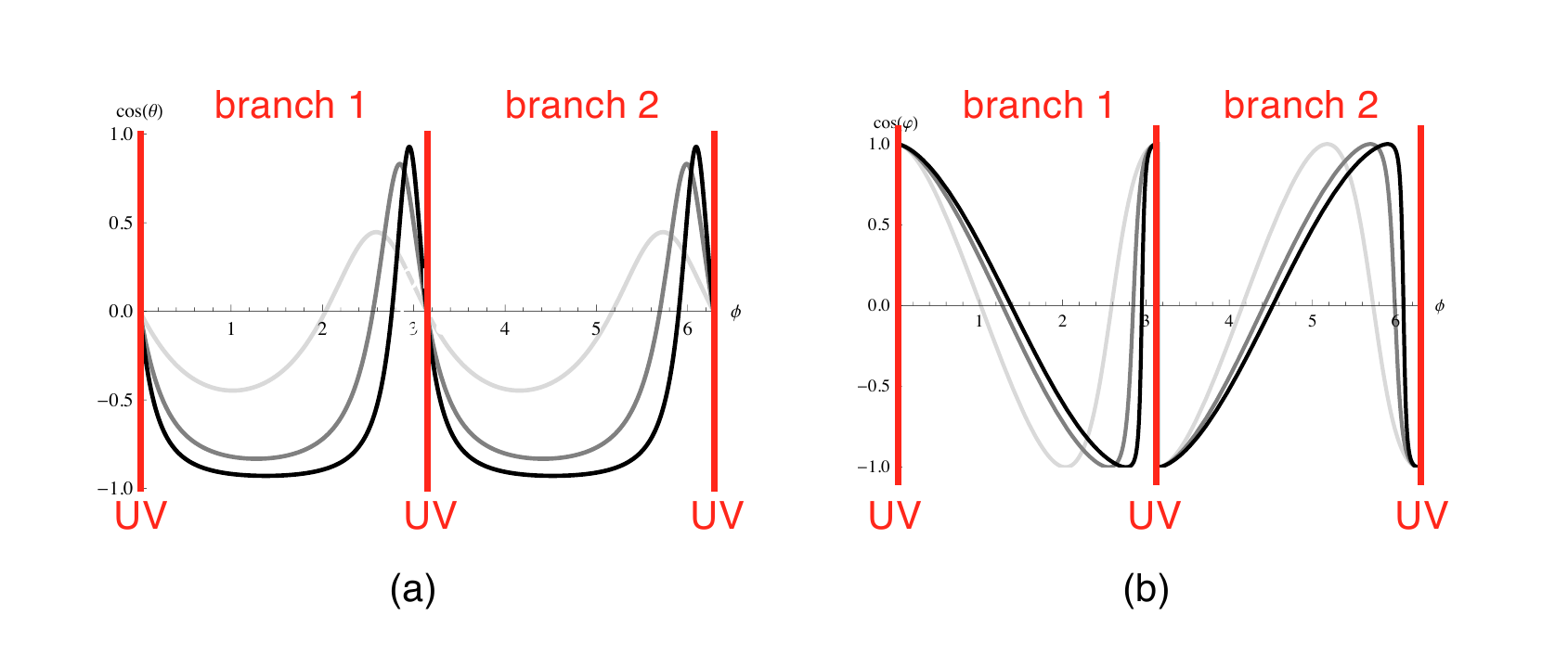} 
	\end{center}
	\caption{\em The probe D3 brane in the (a) $\cos\theta$ and (b) $\cos\varphi$ versus $\phi$ plane; we set $C_2=1$, $C_3=0$, and $C_4=1$, and consider time snapshots $T=1,3,5$ from light gray to black. The vertical bars correspond to the location of the defect in the UV.} \label{fig:fig23-profile} 
\end{figure}

The more interesting case arises from the second possibility $1-g\left(\Phi\right){}^3 g''\left(\Phi\right)=0$. We can solve this equation easily 
\begin{equation}
	g(\Phi )^2=C_2 \Phi^2+2 C_2 C_3 \Phi +C_2^{-1}+C_2 C_3^2 
\end{equation}
where $C_2$ and $C_3$ are constants of integration. We then have 
\begin{equation}
	\varphi = \phi+C_1+\int \frac{d\Phi}{C_2\Phi^2+2\Phi C_2 C_3 + C_2 C_3^2+C_2^{-1}}=\phi+C_1+\tan^{-1} \left[ C_2 \left( C_3+\Phi\right) \right] 
\end{equation}
which we can use in~(\ref{eq:remainingII}) to determine $f(\phi)$ (and hence $\Phi$) 
\begin{equation}
	\frac{f'(\phi )}{f(\phi )}=-\cot \phi\ \ \ ,\ \ \ C_1=\frac{\pi}{2}\Rightarrow f(\phi)= C_4 \csc \phi 
\end{equation}
with another integration constant $C_4$. Note that we have temporarily set $C_1=\pi/2$ to simplify the computations. The final solution for the embedding then takes the form 
\begin{eqnarray}
	&&w^2=C_4^2 \csc ^2(\phi )-2 C_2 C_3 \left(C_4^2 \cot (\phi )+T\right)+C_2 \left(C_4^2 \cot (\phi )+T\right)^2+\frac{1}{C_2} + C_2 C_3^2\nonumber \\
	&&\cos \theta = \frac{2 C_4^2 \csc ^2(\phi )}{C_4^2 \csc ^2(\phi )-2 C_2 C_3 \left(C_4^2 \cot (\phi )+T\right)+C_2 \left(C_4^2 \cot (\phi )+T\right)^2+{C_2}^{-1}+C_2 C_3^2}-1 \nonumber \\
	&&\varphi = \tan ^{-1}\left(C_2 \left(C_3-C_4^2 \cot \phi-T\right)\right)+\phi + \frac{\pi}{2} 
\end{eqnarray}
parameterized by four constants of integration $C_2$, $C_3$, $C_4$, and a shift in $\phi$ due to the Killing vector~(\ref{eq:K4}) (presumably related to $C_1$). This is a dynamic profile which reaches all the way to the UV boundary 
\begin{equation}
	w\rightarrow \infty \Rightarrow \sin\phi\rightarrow 0\ . 
\end{equation}
Figure~\ref{fig:fig22-profile} shows a profile of the probe brane embedding. 
The capping off point in the bulk is time dependent 
\begin{equation}
	\frac{C_2}{C_2 C_4^2+1} T^2=w^2-\frac{1}{C_2}-C_4^2 
\end{equation}
looking very much like the case in scenario I (see Figure~\ref{fig:fig21-profile1}(a)), with 
\begin{equation}
	w\rightarrow \infty \Rightarrow w_{cap}\rightarrow \sqrt{\frac{C_2}{C_2 C_4^2+1}} |T|\ . 
\end{equation}
Hence, the cap never reaches the boundary in finite time. The large $w$ limits of the angular profile functions are 
\begin{equation}
	\cos\theta \rightarrow \frac{1-C_2 C_4^2}{1+C_2 C_4^2}\ \ \ ,\ \ \ \varphi \rightarrow 0 \ \mbox{or}\ \pi\ . 
\end{equation}
Figure~\ref{fig:fig23-profile} shows profiles of these angular directions.
We see that it has two branches: the probe D3 brane folds in the bulk and lands on the boundary on a plane; however, in the $\phi$ direction, it is a stack of two sheets converging on top of each other near the UV boundary. Meanwhile, each branch starts at the equator in $\theta$, moves to the north then south poles, and comes back to the equator; and also makes a full circular trip along $\varphi$. Note that the two branches end at diametrically opposing points along $\varphi$. Figure~\ref{fig:main}(b) in the Introduction shows a cartoon of the setup. Again, the size of the $\phi$ circle shrinks in the UV; hence, to determine whether both branches land at the same point in the ten dimensional space at the defect, we need to consider the T-dual picture. The T-duality being along $\phi$, we can conclude that the defect may be viewed as $D2$ brane probes with maximally $U(2)\times U(2)$ non-Abelian degrees of freedom: we cannot realize $U(4)$ because the two branches are at diametrically opposing points along a large direction $\varphi$ at the defect. We come back to this issue in the Discussion section. For now, we will refer to this configuration as the `figure eight' defect.

The action before subtraction of the boundary terms is 
\begin{equation}
	\mathcal{S} = -i \int dT\,d^2 X d\phi\cot \left(\frac{\theta }{2}\right) \csc \phi \csc (\phi -\varphi )\ . 
\end{equation}
After subtraction, it becomes 
\begin{equation}
	\label{eq:SII} \mathcal{S}\rightarrow -i \int dT\, d^2 X d\phi \cot \left(\frac{\theta }{2}\right) \sin \phi \csc (\phi -\varphi)\ . 
\end{equation}
Substituting the solution into this, we find a constant integral
\begin{equation}
	\mathcal{S} = i C_4 \sqrt{C_2} \int dT\, d^2 X d\phi = 4 \pi i C_4 \sqrt{C_2} T_0 \int d^2 X = 4 \pi i C_4 \sqrt{C_2} T_0 A
\end{equation}
with the $\phi$ integral between $0$ and $2\pi$, the $T$ integral between $-T_0$ to $T_0$, and the regularized area of the operator written as $A$. This gives the vev of the corresponding operator as
\begin{equation}
	\langle \mathcal{O} \rangle \simeq e^{-4 \pi C_4 \sqrt{C_2} T_0 A} = e^{-2 \pi C_4 \sqrt{C_2} \frac{V}{G\Delta^3}}
\end{equation}
where we see again the volume $V=2\, t_0 A$ {\em per puff volume $G\Delta^3$} make an appearance.

Evaluating the energy of the configuration using~(\ref{eq:energy}) (without boundary subtractions), we find 
\begin{equation}
	E=0\ . 
\end{equation}
Again, we expect a simple relation between this energy, the mass dimension of the operator, and the R-charge.

We then computer $Q_4$ using ~(\ref{eq:charge}) and we find 
\begin{equation}
	Q_4= \int d\phi \frac{w^6+1}{4\, w\, \mathcal{L}_{DBI}} \left( w^2 \left(\sin ^4 \theta \csc ^6 \frac{\theta }{2} \csc ^2(\phi -\varphi )+8 (\cos \theta +1) \csc ^2 \phi \right)+16 \left(w^2-1\right) \right)\ . 
\end{equation}
Unfortunately, this expression is again too difficult to simplify further. We can see however that the expression can be non-zero numerically. Given that the state is BPS and the energy vanishes, the mass dimension of the corresponding operator is probably determined by this R-charge.

The last case is scenario IIc of ~(\ref{eq:scenarioIIa})-(\ref{eq:scenarioIIc}) and~(\ref{eq:scenarios}). This case leads to 
\begin{equation}
	(1+ i\,\gamma^{0518})\epsilon_0 = 0\ . 
\end{equation}
One finds that the embedding can be obtained from the current one (scenario IIb) by 
\begin{equation}
	\phi\rightarrow \phi+\frac{\pi}{2}\ \ \ \mbox{and}\ \ \ \varphi\rightarrow \varphi+\frac{\pi}{2}\ . 
\end{equation}
It is the result of using the Killing vector 
\begin{equation}
	K_3-\frac{1}{2} K_4 
\end{equation}
onto scenario IIb. So, the two configurations, scenarios IIb and IIc, are related. But they break different supersymmetries 
\begin{equation}
	(1- i\,\gamma^{0512})\epsilon_0 = 0\ \ \ \mbox{or}\ \ \ (1+ i\,\gamma^{0518})\epsilon_0 = 0\ . 
\end{equation}
Figure~\ref{fig:main}(b) shows the two configurations as solid and dashed profiles. Note however the two configurations are also rotated in the $\varphi$ direction which is not shown in the figure.

\section{Spherical embeddings}

As mentioned previously, PFT preserves spatial rotational symmetry $SO(3)$. The non-local states are expected to carry fractional D3 brane charge as if they are open D3 spherical bubbles. It would then be interesting to realize the defects of the previous sections as spherical formations instead of planar ones: the probe D3 brane would land in the UV on a boundary that is a 2-sphere. To see how this may happen, we would write the $X^1$, $X^2$, and $X^3$ coordinates of the worldvolume in terms of spherical ones $R, \Theta, \varPhi$, leading to the additional Killing equations for $\epsilon_0$
\begin{equation}
	\partial_\Theta \epsilon_0 = \frac{1}{2} \gamma^{56} \epsilon_0\ \ \ ,\ \ \ 
	\partial_\varPhi \epsilon_0 = \frac{1}{2} \cos \Theta \gamma^{67} \epsilon_0 + \frac{1}{2} \sin \Theta \gamma^{57} \epsilon_0\ . 
\end{equation}
Naturally, the spinor $\epsilon_0$ cannot be constant in this coordinate system and rotates as 
\begin{equation}
	\label{eq:rotational} \epsilon_0 = e^{\frac{1}{2} \Theta \gamma^{56}} e^{\frac{1}{2} \Phi \gamma^{67}} \varepsilon_0 
\end{equation}
with now $\varepsilon_0$ being the constant spinor satisfying 
\begin{equation}
	(1-i \gamma^{0567}) \varepsilon_0 = 0 \ \ \ ,\ \ \ (1 -\gamma^{1238}) \varepsilon_0 = 0\ . 
\end{equation}
Note in particular that we still have $[\gamma^{0567},e^{\frac{1}{2} \Theta \gamma^{56}} e^{\frac{1}{2} \Phi \gamma^{67}}]=0$. Throughout the BPS analysis of the probe, we simply need to substitute $X^1\rightarrow R$; the $X^1=\mbox{constant}$ condition arising from the analysis would indicate a defect of finite spherical radius. However, the new Killing spinor $\epsilon_0$ given by~(\ref{eq:rotational}) conflicts with the defect BPS conditions~(\ref{eq:scenarios}). As such, extending our solutions to spherical forms is not possible.

We will next speculate on the reason and remedy for this situation. If we are to realize the non-local states of the PFT as spherical defects, we also naturally need to {\em excise} a ball of D3 brane from the worldvolume before inserting the probe D3 brane on its sphere of a boundary\footnote{This may be viewed as saying that realizing the non-local puff states without taking into account back reaction from the probe may not be suitable. In a sense, the puffs may need the analogue of bubbling geometries for a full description~\cite{Lin:2004nb}\cite{Gomis:2007fi}}. The surgery would create a spherical boundary on the worldvolume. A Killing spinor on a 2-sphere would then instead satisfy the equations~\cite{Lu:1998nu}\cite{Lunin:2006xr}
\begin{equation}
	\nabla_\mu \epsilon_0 = \frac{i}{2} \gamma_\mu \epsilon_0\ .
\end{equation}
This results in a Killing spinor that looks slightly different from~(\ref{eq:rotational})
\begin{equation}
	\label{eq:rotational2} \epsilon_0 = e^{\frac{i}{2} \Theta \gamma^{6}} e^{\frac{1}{2} \Phi \gamma^{67}} \varepsilon_0\ .
\end{equation}
Note that the 5 index corresponds to the radial direction, while the 6 and 7 correspond to $\Theta$ and $\Phi$. We then have removed the $5$ index from the Killing spinor, which renders the new Killing spinor compatible with the probe BPS conditions~(\ref{eq:scenarios}). We believe that this intuitively realizes spherical defects, suggesting that the non-local puffs of the PFT involve a surgical cut of the worldvolume at the location of non-local operator insertion, somewhat akin to similar operations on the string worldsheet when vertex operators are inserted. In this case however, the insertion is a two dimensional closed surface instead of a point.

\section{Discussion}

PFT is interesting because of three main factors: (1) It seems to admit a new elaborate holographic dictionary; (2) Its spectrum realizes non-local states with rich structure reminiscent of surface operators; and (3) It lends itself for real-world physical applications in cosmology. In this work, we have shown that states in PFT carrying twist R-charge can be realized as D3 brane probes in the holographic picture of the PFT. These probes have all the attributes to be interpreted as `volume operators', akin to surface operators in local gauge theories. They end on co-dimension one defects in the PFT while folding  through several branches in the bulk. The result is a non-Abelian character for the defect sigma model, from $U(8)$ to $U(2)\times U(2)\times U(2)\times U(2)$. These are defects of the non-rigid type - with four free parameters - and ordered in the sense that they are similar to Wilson lines as opposed to disordered 't Hooft operators. In principle, it should be straightforward to write a defect theory from a reduction of the appropriate DBI action to the PFT worldvolume. The details of this theory may help identify the degrees of freedom of the PFT. We also alluded to $1/4$ BPS configurations that should be easy to write down. And we have shown that there are no static probe configurations that can admit a volume operator interpretation.

For spherical defects, the setup appears to require surgery on the PFT worldvolume - the excising of balls of the worldvolume and the gluing of the probe on the resulting 2-sphere boundary. The novel holographic dynamics described in~\cite{FreedBrown:2009py}, with operator insertions in the bulk, must play a role in interpreting this D3 brane probe picture. This however requires a proper treatment of spherical configurations - so as to associate the volume of the operator with the location of the insertion in the bulk. It would also be interesting to see whether one can consider bubbling geometries that involve volume operators with non-negligible tension and full back-reaction. In conclusion, we have presented in this work strong evidence that the the isotropic non-local states of the PFT are indeed protrusions from the PFT worldvolume and can be viewed as volume operators.

Several aspects of the discussion raise interesting questions for future directions. It would be useful to understand the role of the time dependence of the configurations from the PFT perspective. Perhaps insertions of two volume operators would generate straddling profiles in the bulk such that the setup remains static. Such a configuration would be needed to compute correlation functions of two volume operators. It would also be interesting to compute correlators of a volume operator with a local operator (along say a treatment similar to~\cite{Drukker:2008wr}), perhaps using a geodesic language with modified boundary conditions instead. All these will help to unravel the structure of PFT, as probed by its non-local states - with the ultimate goal of realizing a computational tractable and complete UV definition of PFT at weak and strong coupling.

\section{Acknowledgments}

I am grateful to Lotte Hollands and John Schwarz for discussions, and the Caltech theory group for hospitality. This work is supported by DOE grant DE-FG02-92ER40701.

\section{Appendices} \label{sec:appendices}

\subsection{$\Delta$'s of equation~(\ref{eq:mainbps})} \label{sub:bpsapp}

In this appendix, we list the explicit forms of the $\Delta$'s appearing in~(\ref{eq:mainbps}). It is convenient to first write these as 
\begin{equation}
	\Delta^{0367} = \cos \frac{\theta}{2} \left(\Delta_3 \cos \phi-\Delta_1 \sin \phi\right)-\sin \frac{\theta}{2} \left(\Delta_2 \sin \phi+\Delta_5 \cos\phi\right) 
\end{equation}
\begin{equation}
	\Delta^{03} = \cos \frac{\theta}{2} \left(\Delta_8 \sin \phi-\Delta_{13} \cos\phi\right)+\sin \frac{\theta}{2} \left(\Delta_{11} \sin \phi-\Delta_{15} \cos\phi\right) 
\end{equation}
\begin{equation}
	\Delta^{58} =-\sin (\phi -\varphi ) \left( \Delta_1 \sin \frac{\theta}{2} -\Delta_2 \cos \frac{\theta}{2} \right)+\cos (\phi -\varphi ) \left(\Delta_3 \sin \frac{\theta}{2}+\Delta_5 \cos \frac{\theta}{2}\right) 
\end{equation}
\begin{equation}
	\Delta^{08} = \sin (\phi -\varphi ) \left( \Delta_8 \sin \frac{\theta}{2} -\Delta_{11} \cos \frac{\theta}{2} \right)+\cos (\phi -\varphi ) \left(\Delta_{15} \cos \frac{\theta}{2}-\Delta_{13} \sin \frac{\theta}{2}\right) 
\end{equation}
\begin{equation}
	\Delta^{15} = -\cos \frac{\theta }{2} \left( \Delta_1 \cos\phi+\Delta_3 \sin \phi \right) -\sin \frac{\theta}{2} \left(\Delta_2 \cos\phi-\Delta_5 \sin \phi\right) 
\end{equation}
\begin{equation}
	\Delta^{01} =-\cos \frac{\theta}{2} \left( \Delta_8 \cos\phi+\Delta_{13} \sin \phi \right)-\sin \frac{\theta}{2} \left(\Delta_{15} \sin \phi+\Delta_{11} \cos\phi\right) 
\end{equation}
\begin{equation}
	\Delta^{0267} = \cos \frac{\theta}{2} \left(\Delta_2 \cos (\phi -\varphi )-\Delta_5 \sin (\phi -\varphi )\right)-\sin \frac{\theta}{2} \left(\Delta_3 \sin (\phi -\varphi )+\Delta_1 \cos (\phi -\varphi )\right) 
\end{equation}
\begin{equation}
	\Delta^{02} = \sin \frac{\theta}{2} \left(\Delta_{13} \sin (\phi -\varphi )+\Delta_8 \cos (\phi -\varphi )\right)-\cos \frac{\theta}{2} \left(\Delta_{15} \sin (\phi -\varphi )+\Delta_{11} \cos (\phi -\varphi )\right) 
\end{equation}
\begin{equation}
	\Delta^{0158} = \left(\Delta_9-\Delta_{10}\right) \cos \theta \cos \varphi+\left(\Delta_7+\Delta_{12}\right) \sin \theta \cos \varphi+\left(\Delta_{14}-\Delta_6\right) \sin \varphi 
\end{equation}
\begin{equation}
	\Delta^{0135} = \left(\Delta_9-\Delta_{10}\right) \sin \theta-\left(\Delta_7+\Delta_{12}\right) \cos \theta 
\end{equation}
\begin{equation}
	\Delta^{0125} = -\left(\Delta_7+\Delta_{12}\right) \sin \theta \sin \varphi-\left(\Delta_9-\Delta_{10}\right) \cos \theta \sin \varphi-\left(\Delta_6-\Delta_{14}\right) \cos \varphi 
\end{equation}
Then, depending on the embedding type - parallel, holographic, or $\phi$ wrapping - we have different expressions

\sechead{Parallel embedding}
\begin{equation}
	 \Delta_1  = -\frac{i w \left(
	\partial_1(\varphi ) (\cos \theta-1)+2 
	\partial_1(\phi )\right)}{2 H}, \nonumber
	 \Delta_2  = -\frac{i w \sin \theta \left(w^2 
	\partial_1(\phi ) 
	\partial_0(\varphi )+
	\partial_1(\varphi ) \left(1-w^2 
	\partial_0(\phi )\right)\right)}{2 H},\nonumber 
\end{equation}
\begin{equation}
	 \Delta_3  = \frac{i \left(
	\partial_1(w) \left(w^2 \left(
	\partial_0(\varphi ) (\cos \theta-1)+2 
	\partial_0(\phi )\right)-2\right)-w^2 
	\partial_0(w) \left(
	\partial_1(\varphi ) (\cos \theta-1)+2 
	\partial_1(\phi )\right)\right)}{2 H},\nonumber 
\end{equation}
\begin{equation}
	 \Delta_4  = \Delta = \frac{i \left(w^2 \left(
	\partial_0(\varphi ) (\cos \theta-1)+2 
	\partial_0(\phi )\right)-2\right)}{2 H^2},\nonumber 
\end{equation}
\begin{equation}
	 \Delta_5  = \frac{i w \left(
	\partial_1(\theta ) \left(w^2 \left(
	\partial_0(\varphi ) (\cos \theta-1)+2 
	\partial_0(\phi )\right)-2\right)-w^2 
	\partial_0(\theta ) \left(
	\partial_1(\varphi ) (\cos \theta-1)+2 
	\partial_1(\phi )\right)\right)}{4 H},\nonumber 
\end{equation}
\begin{equation}
	 \Delta_6  = -\frac{1}{2} i w^2 \sin \theta \left(
	\partial_1(\phi ) 
	\partial_0(\varphi )-
	\partial_0(\phi ) 
	\partial_1(\varphi )\right),\nonumber 
\end{equation}
\begin{equation}
	 \Delta_7  = \frac{1}{2} i w \left(
	\partial_0(w) \left(-
	\partial_1(\varphi ) (\cos \theta-1)-2 
	\partial_1(\phi )\right)+
	\partial_1(w) 
	\partial_0(\varphi ) (\cos \theta-1)+2 
	\partial_1(w) 
	\partial_0(\phi )\right),\nonumber 
\end{equation}
\begin{equation}
	 \Delta_8  = \frac{i w \left(
	\partial_0(\varphi ) (\cos \theta-1)+2 
	\partial_0(\phi )\right)}{2 H},\nonumber 
\end{equation}
\begin{equation}
	 \Delta_9  = \frac{1}{4} i w^2 \left(
	\partial_0(\theta ) \left(-
	\partial_1(\varphi ) (\cos \theta-1)-2 
	\partial_1(\phi )\right)+2 
	\partial_1(\theta ) 
	\partial_0(\phi )+
	\partial_1(\theta ) 
	\partial_0(\varphi ) (\cos \theta-1)\right),\nonumber 
\end{equation}
\begin{equation}
	 \Delta_{10}  = \frac{1}{2} i w \sin \theta \left(
	\partial_1(w) 
	\partial_0(\varphi )-
	\partial_0(w) 
	\partial_1(\varphi )\right)\ \ ,\ \  
	 \Delta_{11}  = \frac{i w 
	\partial_0(\varphi ) \sin \theta}{2 H},\nonumber 
\end{equation}
\begin{equation}
	 \Delta_{12}  = \frac{1}{4} i w^2 \sin \theta \left(
	\partial_1(\theta ) 
	\partial_0(\varphi )-
	\partial_0(\theta ) 
	\partial_1(\varphi )\right),\nonumber 
\end{equation}
\begin{equation}
	 \Delta_{13}  = \frac{i 
	\partial_0(w)}{H}\ \ ,\ \ \Delta_{14}\to \frac{1}{2} i \left(w 
	\partial_0(w) 
	\partial_1(\theta )-w 
	\partial_1(w) 
	\partial_0(\theta )\right)\ \ ,\ \ \Delta_{15}\to -\frac{i w 
	\partial_0(\theta )}{2 H} \nonumber
\end{equation}

\sechead{Holographic embedding}
\begin{equation}
	 \Delta_1= -\frac{i w \left(
	\partial_1(\varphi ) (\cos \theta-1)+2 
	\partial_1(\phi )\right)}{2 H}\ \ ,\ \  
	 \Delta_2= -\frac{i w \sin \theta \left(w^2 
	\partial_1(\phi ) 
	\partial_0(\varphi )+
	\partial_1(\varphi ) \left(1-w^2 
	\partial_0(\phi )\right)\right)}{2 H}, \nonumber 
\end{equation}
\begin{equation}
	 \Delta_3= \frac{i \left(w^2 \left(
	\partial_0(\varphi ) (\cos \theta-1)+2 
	\partial_0(\phi )\right)-2\right)}{2 H}, \nonumber 
\end{equation}
\begin{equation}
	 \Delta_4=\Delta = \frac{i \left(
	\partial_1(R) \left(w^2 \left(
	\partial_0(\varphi ) (\cos \theta-1)+2 
	\partial_0(\phi )\right)-2\right)-w^2 
	\partial_0(R) \left(
	\partial_1(\varphi ) (\cos \theta-1)+2 
	\partial_1(\phi )\right)\right)}{2 H^2}, \nonumber 
\end{equation}
\begin{equation}
	 \Delta_5= \frac{i w \left(
	\partial_1(\theta ) \left(w^2 \left(
	\partial_0(\varphi ) (\cos \theta-1)+2 
	\partial_0(\phi )\right)-2\right)-w^2 
	\partial_0(\theta ) \left(
	\partial_1(\varphi ) (\cos \theta-1)+2 
	\partial_1(\phi )\right)\right)}{4 H}, \nonumber 
\end{equation}
\begin{equation}
	 \Delta_6= -\frac{1}{2} i w^2 \sin \theta \left(
	\partial_1(\phi ) 
	\partial_0(\varphi )-
	\partial_0(\phi ) 
	\partial_1(\varphi )\right)\ \ ,\ \ 
	 \Delta_7= \frac{1}{2} i w \left(
	\partial_0(\varphi ) (\cos \theta-1)+2 
	\partial_0(\phi )\right), \nonumber 
\end{equation}
\begin{equation}
	 \Delta_8= \frac{i w \left(
	\partial_0(R) \left(-
	\partial_1(\varphi ) (\cos \theta-1)-2 
	\partial_1(\phi )\right)+
	\partial_1(R) 
	\partial_0(\varphi ) (\cos \theta-1)+2 
	\partial_1(R) 
	\partial_0(\phi )\right)}{2 H}, \nonumber 
\end{equation}
\begin{equation}
	 \Delta_9= \frac{1}{4} i w^2 \left(
	\partial_0(\theta ) \left(-
	\partial_1(\varphi ) (\cos \theta-1)-2 
	\partial_1(\phi )\right)+2 
	\partial_1(\theta ) 
	\partial_0(\phi )+
	\partial_1(\theta ) 
	\partial_0(\varphi ) (\cos \theta-1)\right), \nonumber 
\end{equation}
\begin{equation}
	 \Delta_{10}= \frac{1}{2} i w 
	\partial_0(\varphi ) \sin \theta\ \ ,\ \ 
	 \Delta_{11}= \frac{i w \sin \theta \left(
	\partial_1(R) 
	\partial_0(\varphi )-
	\partial_0(R) 
	\partial_1(\varphi )\right)}{2 H}, \nonumber 
\end{equation}
\begin{equation}
	 \Delta_{12}= \frac{1}{4} i w^2 \sin \theta \left(
	\partial_1(\theta ) 
	\partial_0(\varphi )-
	\partial_0(\theta ) 
	\partial_1(\varphi )\right), \nonumber 
\end{equation}
\begin{equation}
	 \Delta_{13}= -\frac{i 
	\partial_0(R)}{H}\ \ ,\ \ \Delta_{14}= -\frac{1}{2} i w 
	\partial_0(\theta )\ \ ,\ \ \Delta_{15}= \frac{i \left(w 
	\partial_0(R) 
	\partial_1(\theta )-w 
	\partial_1(R) 
	\partial_0(\theta )\right)}{2 H} 
\end{equation}

\sechead{$\phi$ wrapping}
\begin{equation}
	 \Delta_1= -\frac{i w \left(
	\partial_1(\varphi ) (\cos \theta-1)+2\right)}{2 H} \ \ ,\ \  
	 \Delta_2= -\frac{i w \sin \theta \left(w^2 
	\partial_0(\varphi )+
	\partial_1(\varphi )\right)}{2 H}, \nonumber 
\end{equation}
\begin{equation}
	 \Delta_3= -\frac{i \left(2 
	\partial_1(w) \left(w^2 
	\partial_0(\varphi ) \sin ^2\left(\frac{\theta }{2}\right)+1\right)+w^2 
	\partial_0(w) \left(
	\partial_1(\varphi ) (\cos \theta-1)+2\right)\right)}{2 H}, \nonumber 
\end{equation}
\begin{equation}
	 \Delta_4=\Delta -\frac{i \left(2 
	\partial_1(R) \left(w^2 
	\partial_0(\varphi ) \sin ^2\left(\frac{\theta }{2}\right)+1\right)+w^2 
	\partial_0(R) \left(
	\partial_1(\varphi ) (\cos \theta-1)+2\right)\right)}{2 H^2}, \nonumber 
\end{equation}
\begin{equation}
	 \Delta_5= -\frac{i w \left(2 
	\partial_1(\theta ) \left(w^2 
	\partial_0(\varphi ) \sin ^2(\theta/2)+1\right)+w^2 
	\partial_0(\theta ) \left(
	\partial_1(\varphi ) (\cos \theta-1)+2\right)\right)}{4 H}, \nonumber 
\end{equation}
\begin{equation}
	 \Delta_6= -\frac{1}{2} i w^2 
	\partial_0(\varphi ) \sin \theta, \nonumber 
\end{equation}
\begin{equation}
	 \Delta_7= -\frac{1}{2} i w \left(
	\partial_1(w) 
	\partial_0(\varphi ) (1-\cos \theta)+
	\partial_0(w) \left(
	\partial_1(\varphi ) (\cos \theta-1)+2\right)\right), \nonumber 
\end{equation}
\begin{equation}
	 \Delta_8= -\frac{i w \left(
	\partial_1(R) 
	\partial_0(\varphi ) (1-\cos \theta)+
	\partial_0(R) \left(
	\partial_1(\varphi ) (\cos \theta-1)+2\right)\right)}{2 H}, \nonumber 
\end{equation}
\begin{equation}
	 \Delta_9= -\frac{1}{4} i w^2 \left(
	\partial_1(\theta ) 
	\partial_0(\varphi ) (1-\cos \theta)+
	\partial_0(\theta ) \left(
	\partial_1(\varphi ) (\cos \theta-1)+2\right)\right), \nonumber 
\end{equation}
\begin{equation}
	 \Delta_{10}= \frac{1}{2} i w \sin \theta \left(
	\partial_1(w) 
	\partial_0(\varphi )-
	\partial_0(w) 
	\partial_1(\varphi )\right)\ \ ,\ \  
	 \Delta_{11}= \frac{i w \sin \theta \left(
	\partial_1(R) 
	\partial_0(\varphi )-
	\partial_0(R) 
	\partial_1(\varphi )\right)}{2 H}, \nonumber 
\end{equation}
\begin{equation}
	 \Delta_{12}= \frac{1}{4} i w^2 \sin \theta \left(
	\partial_1(\theta ) 
	\partial_0(\varphi )-
	\partial_0(\theta ) 
	\partial_1(\varphi )\right)\ \ , \ \ 
	 \Delta_{13}= \frac{i \left(
	\partial_1(R) 
	\partial_0(w)-
	\partial_0(R) 
	\partial_1(w)\right)}{H}, \nonumber 
\end{equation}
\begin{equation}
	 \Delta_{14}= \frac{1}{2} i \left(w 
	\partial_0(w) 
	\partial_1(\theta )-w 
	\partial_1(w) 
	\partial_0(\theta )\right)\ \ ,\ \ 
	 \Delta_{15}= \frac{i \left(w 
	\partial_0(R) 
	\partial_1(\theta )-w 
	\partial_1(R) 
	\partial_0(\theta )\right)}{2 H} 
\end{equation}

\bibliographystyle{utphys}


\providecommand{\href}[2]{#2}\begingroup\raggedright\endgroup

\end{document}